\documentclass[twocolumn]{aastex701}
\usepackage{subcaption}
\hypersetup{linkcolor=red, urlcolor=magenta}

\begin{document}
\received{April 06, 2026}
\revised{May 22, 2026}
\accepted{May 27, 2026}
\shorttitle{Lowest of the low: SN~2024abfl}
\shortauthors{Teja et al.}

\title{Subluminous Type IIP SN~2024abfl as a Result of a Significantly Low-energy Fe-core Collapse }

\correspondingauthor{Rishabh Singh Teja}

\author[0000-0002-0525-0872]{Rishabh Singh Teja}
\affil{Tsung-Dao Lee Institute, Shanghai Jiao Tong University, 1 Lisuo Road, Shanghai 201210, People’s Republic of China}
\email[show]{rsteja@sjtu.edu.cn, rsteja001@gmail.com}

\author[0000-0002-6688-0800]{D. K. Sahu}
\affil{Indian Institute of Astrophysics, II Block, Koramangala, Bengaluru-560034, Karnataka, India}
\email{dks@iiap.res.in}

\author[0000-0003-3533-7183]{G. C. Anupama}
\affil{Indian Institute of Astrophysics, II Block, Koramangala, Bengaluru-560034, Karnataka, India}
\email{gca@iiap.res.in}

\author[0000-0003-2091-622X]{Avinash Singh}
\affiliation{Oskar Klein Centre, Department of Astronomy, Stockholm University, Albanova University Centre, SE-106 91 Stockholm, Sweden}
\email{avinash.singh@gmail.com}

\author[0009-0008-9062-1455]{Amrit Dutta}
\affiliation{Indian Institute of Astrophysics, II Block, Koramangala, Bengaluru-560034, Karnataka, India}
\affiliation{Pondicherry University, R.V. Nagar, Kalapet, 605014, Puducherry, India}
\email{amrit.dutta@iiap.res.in}

\author[0009-0009-4872-1134]{Gitika Rameshan}
\affiliation{Indian Institute of Astrophysics, II Block, Koramangala, Bengaluru-560034, Karnataka, India}
\affiliation{Academy of Scientific and Innovative Research (AcSIR), Ghaziabad, Uttar Pradesh, 201002, India}
\email{gitika.rameshan@iiap.res.in}

\author[0009-0007-9727-7792]{Hrishav Das}
\affiliation{Indian Institute of Astrophysics, II Block, Koramangala, Bengaluru-560034, Karnataka, India}
\affiliation{Pondicherry University, R.V. Nagar, Kalapet, 605014, Puducherry, India}
\email{hrishav.das@iiap.res.in}

\author[0000-0001-6099-9539]{Koji S Kawabata}
\affiliation{Hiroshima Astrophysical Science Center, Hiroshima University, Higashi-Hiroshima, Hiroshima 739-8526, Japan}
\email{kawabtkj@hiroshima-u.ac.jp}

\author[0000-0001-6706-2749]{Mridweeka Singh}
\affiliation{Indian Institute of Astrophysics, II Block, Koramangala, Bengaluru-560034, Karnataka, India}
\email{mridweeka.singh@iiap.res.in}

\author[0000-0002-6112-7609]{Varun Bhalerao}
\affiliation{Department of Physics, Indian Institute of Technology Bombay, Powai, Mumbai 400076}
\email{varunb@iitb.ac.in}

\begin{abstract}

We present extensive, well-sampled multiwavelength photometric and low-resolution optical spectroscopic observations of the low-luminosity Type IIP supernova SN~2024abfl. SN~2024abfl is found to be at the faintest end of Type IIP supernovae with unprecedented flat (0.1~mag/ 100 day) plateau evolution and a mid-plateau absolute magnitude of $\rm M_V\approx-13.8~mag$, placing it among one of the faintest Type IIP supernovae discovered to date. SN~2024abfl is adjacent to SN~2018zd in the same host NGC~2146. Using various SN distance measurement probes, we provide independent estimates of the debated distance to the host NGC~2146 (7-9~Mpc). Spectral evolution of  SN~2024abfl is found to be similar to other SNe spectra of this subclass but with very narrow line profiles,  indicating moderately low expansion velocities of the ejecta. Detailed 1-D hydrodynamical modeling suggests a compact progenitor with an upper limit of 10~M$_\odot$, fairly consistent with the directly detected progenitor estimates. It exploded with very low-energy 0.05 foe or less with a very low nickel mass of 0.003~M$_\odot$,  consistent with the observed parameters. These parameters provide important constraints on the nature of low-energy core-collapse explosions. We discuss possible progenitor scenarios and compare SN~2024abfl with other low-luminosity Type IIP supernovae.

\end{abstract}


\newpage
\section{Introduction} 

The evolution of massive stars beyond 8-10~M$_\odot$ primarily ends in violent explosions known as core-collapse supernovae (CCSNe), leaving behind compact remnants such as neutron stars or black holes. These are actively pursued to understand the chemical evolution of galaxies, the formation channels for various heavy elements, star formation to test stellar evolution theories, and the formation of dust \citep[][and references therein]{Alsabti2017}. Broadly, CCSNe are categorized as hydrogen-rich (Type II) and hydrogen-poor (Type Ib/c) based on the presence/absence of hydrogen Balmer lines in their optical spectrum \citep{1997Filippenko, 2017hsn..book..769S}. Type II supernovae (SNe) arise from progenitors that retain a substantial hydrogen envelope at the time of explosion and exhibit a wide range of properties in their light curves and spectra. Type II SNe have additional distinctions based on specific spectral features. These SNe also have a unique historical classification based on their light-curve evolution \citep{1979Barbon}, with SNe showing a `plateau' phase in their luminosities referred to as Type IIP SNe, whereas those showing a light-curve decline `linearly' are termed as Type IIL SNe \citep{2012ApJ...756L..30A, 2014MNRAS.445..554Fd3}. Recently, with enough data and detailed modeling, this differentiation seems to be fading away in favor of a continuous population of these events \citep{2014Anderson, 2026arXiv260215788C}. Apart from this, their light curves show varied plateau lengths, luminosities, decline rates, and rise times, among other characteristics \citep{1979Barbon, 2014Anderson, Faran2014_2005cs, 2019MNRAS.488.4239P}. 

While the majority of Type IIP SNe show peak magnitudes around  $M_V\sim-16$~mag, a clear subgroup had emerged over the past couple of decades that shows fainter magnitudes systematically \citep{2014Anderson}. These are now classified into a distinct subgroup of intermediate- to low-luminosity Type IIP (LLIIP) supernovae \citep{2026PASP..138b4204D}.  These events are further characterized by low explosion energies ($\rm E_{exp}\sim10^{50}~erg$), low nickel masses ($\rm M_{Ni}\sim10^{-2}-10^{-3}~M_\odot$), extended plateau lengths and slow expansion velocities. Some of the well-studied examples from this subclass include SN~1997D \citep{1998ApJ...498L.129T}, SN~2005cs \citep{Pastorello2006_2005cs, 20062005cs}, SN~2009md \citep{2011MNRAS.417.1417F}, and SN~2022acko \citep{2023ApJ...953L..18B}.

Even with numerous detailed studies over the past few decades, the understanding of the progenitors of low-luminosity SNe remains inconclusive. Several pathways are proposed as plausible origins of these types of supernovae (SNe), both from low-mass \citep[$\rm9-12~M_\odot$][]{2006ApJ...641.1060L, 2021MNRAS.503..797K,2022ApJ...934...67B, 2025A&A...694A.260D, 2025arXiv251107540V} which also includes the possibility of electron-capture SNe \citep{2021MNRAS.503..797K} to moderate-mass progenitors \citep[$\sim15~M_\odot$][]{2013A&A...555A.145U, 2014LLIIP, 2017MNRAS.464.3013P, 2022MNRAS.513.4983V} going through Fe-core collapse with non-negligible ejecta falling onto the core (fallback) and at times even massive-star progenitors \citep[$>\rm17-18~M_\odot$][]{1998ApJ...498L.129T, 2001MNRAS.322..361B, 2003MNRAS.338..711Z, 2008A&A...491..507U} with significant fallback, that could result in low-luminosity Type II SNe. In some cases, massive stars with significant mass loss prior to collapse could also produce these faint supernovae \citep{2024ApJ...974...44T}. Constraints on these progenitor masses, derived from various techniques, remain in tension. Direct imaging of the LLIIP sites typically indicates less massive RSGs \citep{2009ARA&A..47...63S, 2017RSPTA.37560277V, 2025ApJ...982L..55L} whereas analytical and hydrodynamical models yield a multitude of progenitor masses. More recent efforts favor low-energy explosions of less massive RSGs, reducing the tension with directly detected progenitors. Nevertheless, low-luminosity SNe provide a great opportunity to test the extremes of stellar evolution, and studying another low-luminosity SN increments the limited number of such observed SNe. 

SN~2024abfl was discovered on 2024-11-15 (JD 2460630.08) by an amateur astronomer, Koichi Itagaki \citep{2024TNSTR4506....1I}, at 17.5 mag in \textit{Clear-}filter. We classified SN~2024abfl as a young Type II supernova \citep{2024TNSCR4515....1D} based on the presence of weak hydrogen lines and initiated regular follow-up observations in both spectroscopy and photometry. Several retrospective non-detections were later reported, with 18.29~mag (\textit{Clear-}filter) on JD~2460630.08 being the closest to the explosion epoch. It was, however, not deep enough to provide a conclusive estimate of the explosion epoch. Another non-detection was reported for JD~2460628.09 with 19~mag in \textit{Clear} filter. We use the mean of this and the discovery date to estimate the explosion epoch as $\approx$2460629$\pm$1, which we use as our reference explosion epoch throughout the analysis. Being a nearby SN and in close proximity ($6''$) to the renowned SN~2018zd, SN~2024abfl was widely followed. \citet{2025ApJ...982L..55L} searched for its candidate progenitor star using archival HST images and estimated a moderately reddened RSG progenitor of 9-12~$M_\odot$ from a distance range of 12.6 - 21.7 Mpc. Furthermore, \citet{2026arXiv260102638G, 2026arXiv260401806L} analyze and discuss its photometric and spectroscopic evolution and obtain various observational properties and progenitor estimates. In this work, we extend these studies by providing extensive photometric and spectroscopic monitoring, an independent distance measurement of the host galaxy, and an estimate of the properties of the progenitor star using hydrodynamical modeling.    

The structure of this paper is as follows: firstly, we describe the multi-band photometric and spectroscopic observations for this object in Section~\ref{sec:obs}. In Section~\ref{sec:distance}, we perform an independent distance analysis to constrain the distance to the host. Section~\ref{sec:lightcurve} gives the light curve evolution, comparisons, and nickel mass estimates. In Section~\ref{sec:spec}, we discuss its spectral evolution while comparing with other well-studied low-luminosity objects. We then attempt to model its explosion and progenitor parameters using 1D MESA+STELLA hydrodynamical modeling in Section~\ref{sec:origins}. We end this work by providing a brief summary in \ref{sec:end}.

\section{Observations}
\label{sec:obs}
\begin{figure*}[htb!]
    \centering
   
    \resizebox{\hsize}{!}{\includegraphics{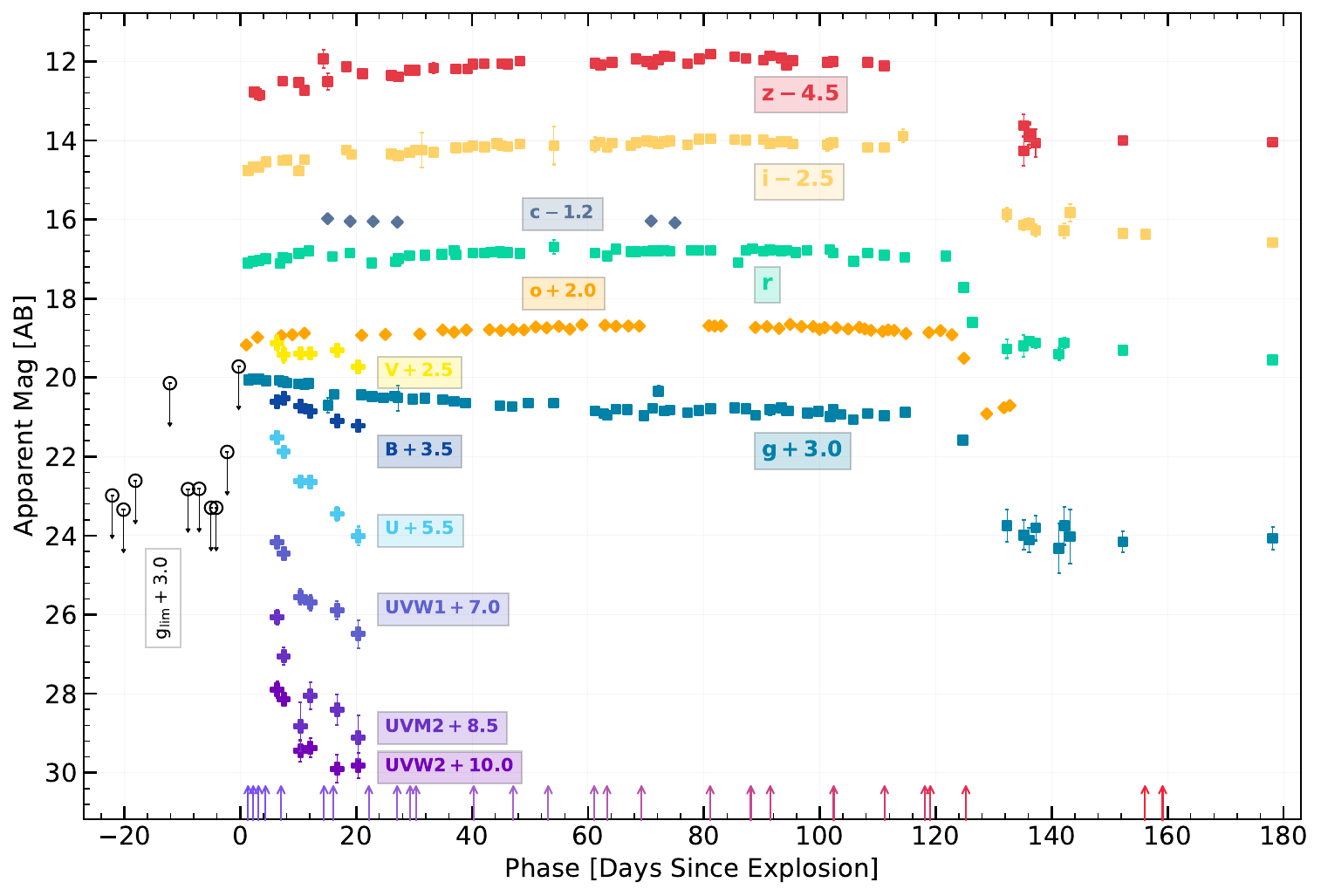}}
        \caption{Light curve evolution of SN~2024abfl for various filters from GIT and HCT is shown. The light curves also include data from ZTF, ATLAS and Swift/UVOT. The constants added to the individual light curves are for visual clarity. We also mark the spectra epochs with upward arrows at the corresponding phases. (All magnitudes are in the AB system) }
    \label{fig:lightcurves} 
     \end{figure*}

 After the discovery, we started continuous photometric and spectroscopic monitoring of SN~2024abfl. High cadence photometry was obtained using 
 0.70-m GROWTH-India Telescope \citep[GIT,][]{2022AJ....164...90K} in SDSS-$g'r'i'z'$ filters. We have also obtained photometry from the 2.0-m Himalayan Chandra Telescope \citep[HCT,][]{2014Prabhu} in the SDSS-$g'r'i'z'$ filter set,  at several epochs, particularly during the post-plateau phase when the object became very faint. We also supplement our photometry with the publicly available data in ZTF  \citep{2019ZTF}-$g$ and -$r$ bands via ALeRCE \citep{2021AJ....161..242F}, and in $c$ and $o$ bands from Asteroid Terrestrial-impact Last Alert System \citep[ATLAS,][]{2018PASP..130f4505T, 2020PASP..132h5002S} via forced photometry server \citep{2021TNSAN...7....1S}).  
 We also performed NUV photometry on the archival data obtained from UltraViolet/Optical Telescope \citep[][$UVOT$]{2005roming} on board Neil Gehrels {\it Swift} Observatory \citep{2004ApJ...611.1005G} for all the available epochs. Standard procedures have been used to carry out the data reduction process using IRAF and Python. SN~2024abfl, located in an isolated, source-free region, shows no contamination from the host, and hence we adopted point-spread-function photometry. The observed light curve of SN~2024abfl in different bands is shown in Figure 1, and the associated data are given in Appendix~\ref{app:data}.
 
 Optical spectra were obtained with the HFOSC instrument available on HCT using 167l slit (1.$\arcsec$92 width and 11$\arcmin$ length). The spectra observed with grisms Gr7 and Gr8 were combined to obtain spectra covering a wavelength range of 4000 to 9000~\AA. The optical spectra were obtained during 1.3 - 160~d post-explosion. The epoch of spectroscopic observations are  marked at the bottom of Figure 1 and a spectroscopy log is also shown in Table~\ref{tab:HCTspec}. Beyond 160~d, in the late nebular phase, the SN faded considerably, and spectroscopy with HCT was not feasible.  The observed spectra were reduced in a standard manner using various tasks available in IRAF. Detailed photometric and spectroscopic reduction procedures are described in details in \citet{2023ApJ...954..155T} and \citet{2024ApJ...974...44T}. 

\begin{table}[hbt!]
\centering
\caption{Log of spectroscopic observations of SN~2024abfl.} 
\begin{tabular}{cccc} \hline
   Date          & JD         &Phase    & Range       \\        
(yyyy-mm-dd)     & (2460000+) &(d)      & (\AA)       \\        
\hline
2024-11-15        & 630.3   & $+$1.3      &  3800--9000 \\     
2024-11-16        & 631.2   & $+$2.2      &  3800--9000 \\     
2024-11-17       & 632.1   & $+$3.1      &  3800--9000 \\     
2024-11-18        & 633.3   & $+$4.3     &  3800--9000 \\     
2024-11-21       & 636.0   & $+$7.0     &  5400--7900 \\     
2024-11-28        & 643.4   & $+$14.4     &  3800--7900 \\     
2024-11-30        & 645.0   & $+$16.0     &  3800--9000 \\     
2024-12-06        & 651.1   & $+$22.1     &  3800--9000 \\     
2024-12-11       & 656.0   & $+$27.0     &  3800--7200 \\     
2024-12-13       & 658.3   & $+$29.3     &  3800--9000 \\     
2024-12-14        & 659.3   & $+$30.3     &  3800--9000 \\       
2024-12-24      & 669.2   & $+$40.2     &  5100--7200 \\     
2024-12-31        & 676.1  & $+$47.1     &  3800--9000 \\     
2025-01-06        & 682.1   & $+$53.1     &  3800--9000 \\     
2025-01-14        & 690.0   & $+$61.0    &  3800--9000 \\      
2025-01-16        & 692.2   & $+$63.2     &  3800--9000 \\             
2025-01-22        & 698.2   & $+$69.0     &  3800--9000 \\     
2025-02-03       & 710.0   & $+$81.0     &  3800--9000 \\     
2025-02-10        & 717.0   & $+$88.0     &  3800--9000 \\     
2025-02-13        & 720.4   & $+$91.4     &  3800--7900 \\     
2025-02-14        & 731.3   & $+$102.3     &  3800--9000 \\     
2025-03-05       & 740.1   & $+$111.1    &  3800--9000 \\       
2025-03-12        & 747.1   & $+$118.1     &  3800--9000 \\       
2025-03-13        & 748.0   & $+$119.0    &  5100--9000 \\     
2025-03-19        & 754.2   & $+$125.2    &  4100--9000 \\     
2025-04-19         & 785.1   & $+$156.1    &  5100--9000 \\             
2025-04-22        & 788.1  & $+$159.1   &  4100--9000 \\         
\hline
\end{tabular} \\
$^{\ast}$ With reference to the explosion date (JD~2460629). 
\label{tab:HCTspec}
\end{table}

\section{Distance to the Host}
\label{sec:distance}
The distance to NGC~2146 remains uncertain, with available distances in the literature ranging from $\sim$ 13 to 22 Mpc \citep{2021arXiv210912943C}. The inferred brightness and explosion parameters of SN~2024abfl are sensitive to the adopted distance. Hence, it is crucial to constrain the distance of the host galaxy. \citet{2026arXiv260102638G} estimated a distance of $\rm 14.42\pm1.87~Mpc$ from SN~2024abfl data by applying the Standardized Candle Method \citep[SCM,][]{Hamuy2002}. The SCM technique, however, is calibrated for SNe with a typical plateau duration of 100~days, and the calibration is presented at the mid-plateau epoch of 50~days. Moreover, around $\rm+50~d$ we find that \ion{Fe}{2}~$\rm \lambda$~5169 has a blended line profile in our spectra and is prone to large uncertainties in measured velocities. Other \ion{Fe}{2} lines that we could use also show a discontinuity in velocity evolution around a similar phase. Therefore, we measure velocity from  $\rm+47.1~d$ spectra for SCM measurements. Even if we do not have velocity at 50~d, the expansion velocity continues to decline, and this could be considered as an upper limit on the SCM distance. The \ion{Fe}{2} velocity was measured using minima of the absorption feature. Using Eqs~(2) and (4) from \citet{2003astro.ph..9122H}, we can estimate the SCM distance to the host for \textit{V} and \textit{I} data, respectively. Assuming $\rm H_0=70~km~s^{-1}~Mpc^{-1}$, we get $\rm D_L\approx9.09\pm3.10~and~8.1\pm2.7~Mpc$, for \textit{V} and \textit{I} relations respectively. Another calibration exists for SCM using the V-I color without color corrections \citep{2006ApJ...645..841N}, rather than the individual magnitudes with color corrections. Using this, we obtained $D_L\approx 13.67^{+2.84}_{-2.35}~Mpc$. The error is dominated by the velocity error and does not include the intrinsic error in the correlation equations. The distance estimates using the $V$ and $I$ magnitudes are similar to the values obtained by \citet{2026arXiv260401806L}, whereas the color term values are very similar to the values obtained by \citet{2026arXiv260102638G}, who used a higher \ion{Fe}{2} velocity ($\rm 1500~km~s^{-1}$). Since we find discrepant estimates using SCM, we employ other methods to estimate the distance. 

Another reliable method, the expanding photosphere method (EPM), can also be used to ascertain the SN distance. The Type IIP supernova SN~2018zd occurred in the same host at a very close-by location \citep{2020MNRAS.498...84Z, 2021NatAs...5..903H, 2021arXiv210912943C}. Utilizing the SCM, \citet{2021NatAs...5..903H} estimated the distance to SN ~2018zd to be $\rm 9.6\pm1~Mpc$. The conspicuous presence of CSM interaction in SN~2018zd light curve, however, prevented them from using EPM. In addition to SCM, the EPM can also be used to impose further constraints on the SN~2024abfl distance. We note that there are indications of the presence of CSM signatures very early on in the spectra \citep{2026arXiv260102638G, 2026arXiv260204309C}, which subside within a few days and thus do not influence our calculations. We proceed to apply EPM to estimate the distance to SN~2024abfl.

\begin{figure}[hbt!]
	 \resizebox{\hsize}{!}{\includegraphics{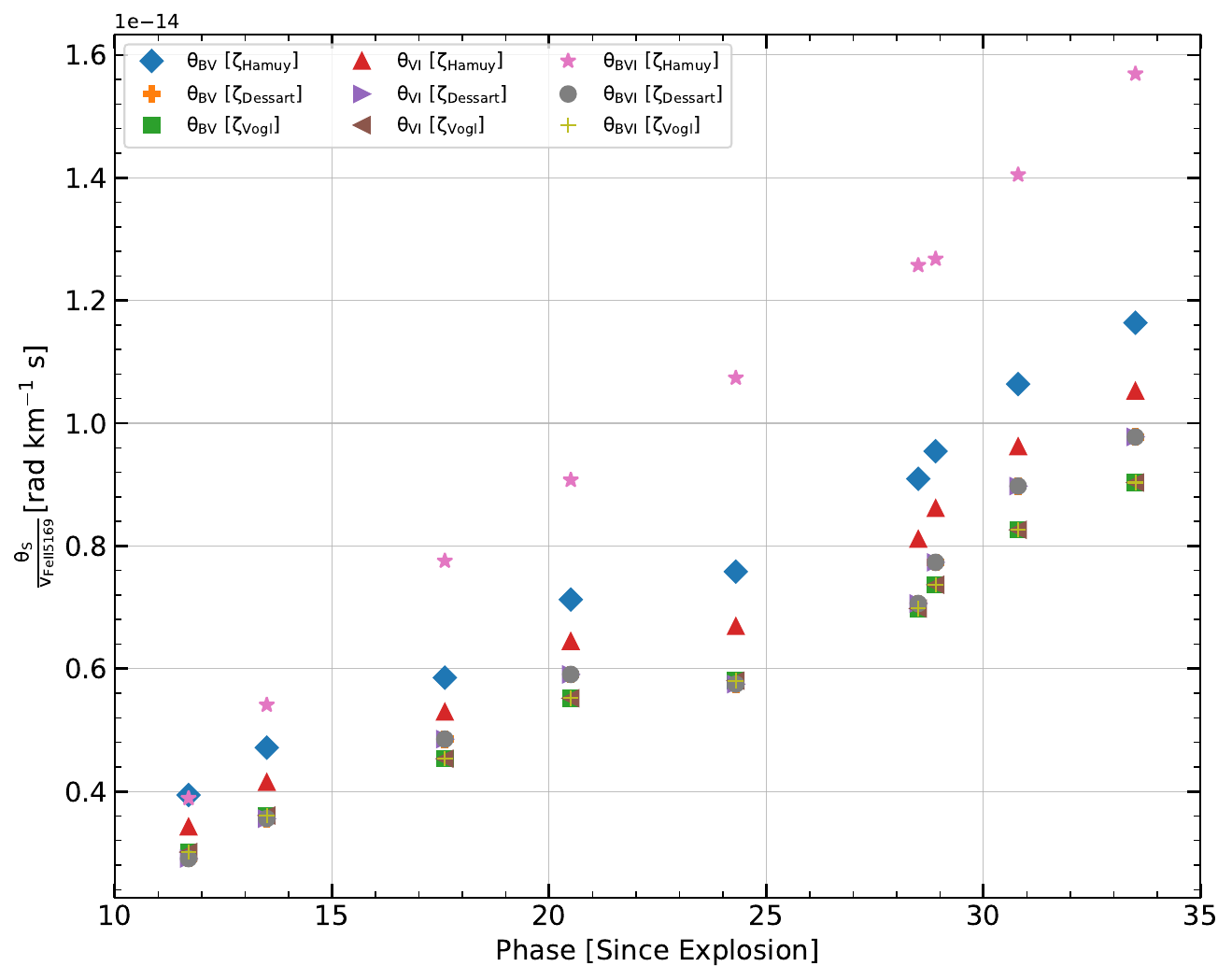}}
    \caption{Plot of $ \frac{\theta_i}{v_i}~vs~(t_i - t_0)$ for various filter sets and different dilution factors from \citet{hamuy2001_1999em}, \citet{DH2005_EPMdilution} and \citet{Vogl2019_EPMDilution}. We estimated slopes (then distances) by applying linear fits to the data points.}
    \label{fig:EPM_values}
    
\end{figure}

The implementation of EPM was followed as per the details given in \citet{hamuy2001_1999em} and \citet{DH2005_EPMdilution}. This formalism involves the measurements of two radii associated with SN: i) a photometric angular radius ($\theta$)  and ii) a spectroscopic physical
radius (R). With the aid of these two radii, the distance to SN could be derived. The angular radius ($\theta$) is given as:
$$\rm \theta = \frac{R}{D} = \sqrt{\frac{f_\lambda}{\pi B_\lambda (T) 10^{-0.4A(\lambda)}\zeta_\lambda^2 }}, $$ where, D is the distance to the SN. $B_\lambda(T)$ is the Planck function at the color temperature of the diluted blackbody radiation, $f_\lambda$ is the apparent flux density, A($\lambda$) is the dust extinction, and $\zeta_\lambda$ is the dilution factor to account for the deviation from a black body \citep{hamuy2001_1999em}. The above equation could be transformed in terms of apparent magnitudes ($m_\lambda$) for multiband photometry as:
$$\rm m_\lambda= -5 log(\zeta_\lambda) - 5 log(\theta) + A_\lambda + b_\lambda(T) ,$$
Now, for different filter sets (S), the above equation was minimized, with $b_\lambda$ values taken from \citet{hamuy2001_1999em}, and dilution factors were taken from three different works, viz. \citet{hamuy2001_1999em}, \citet{DH2005_EPMdilution} and \citet{Vogl2019_EPMDilution}. The quantity that was minimized is as follows:
$$\rm \varepsilon = \sum_{\lambda \epsilon S} \left[ m_\lambda  + log(\theta_S \zeta_S) - A_\lambda - b_\lambda(T_S) \right]^2 $$

Finally, using the expansion velocity ($v$) measured from spectra could be used in the following equation:
 $$ \frac{\theta_i}{v_i} \approx \frac{(t_i - t_0)}{D},$$ where subscript i implies values available for different epochs. A straight line could be fit over multiple epochs, and the resulting slope is used to estimate the distance (Figure~\ref{fig:EPM_values}). We used \ion{Fe}{2}~$\lambda 5169$ velocities estimated from spectra (without redshift correction) and interpolated these with a power law to obtain velocities at the observed photometric epochs ($\rm v_i$).
In these studies, dilution factors required to estimate distance are based on the Bessell-$BVRI$ filters; hence, we converted $griz$ magnitudes to $BVI$ magnitudes using transformation relations from \citep{2005AN....326..657J}. For each dilution factor and filter set, we obtained several distances (Figure~\ref{fig:EPM_values}). The mean distances for all filter sets corresponding to different dilution factors are $7.02\pm1.78$~Mpc, $8.79\pm1.89$~Mpc, and $9.25\pm2.29$~Mpc for \citet{hamuy2001_1999em}, \citet{DH2005_EPMdilution}, and \citet{Vogl2019_EPMDilution}, respectively. The derived EPM distances range from 7 to 9.3 Mpc, depending on the adopted dilution factors, and are broadly consistent with the value reported in \citet{2021NatAs...5..903H}, though with a slightly higher error. Hence, we use distance and other host properties, such as extinction and metallicity, from the values obtained in \citet{2021NatAs...5..903H}. The total extinction includes Milky-Way line of sight extinction obtained through IRSA dust maps \citep{2011ApJ...737..103Sirsa} and an equivalent host extinction as estimated for SN~2018zd \citep{2021NatAs...5..903H} assuming $\rm R_V=3.1$ \citep{Cardelli}. \citet{2026arXiv260102638G} and \citet{2026arXiv260204309C} also estimated similar extinction values within reported error using \ion{Na}{1D} doublet from their respective spectra. In further analysis, we use $\rm D_L=9.6\pm 1.0~Mpc,$, and  $\rm A_V\approx 0.53~mag$.

\section{Light Curve}
\label{sec:lightcurve}
  \begin{figure}[hbt!]
   \resizebox{\hsize}{!}{\includegraphics{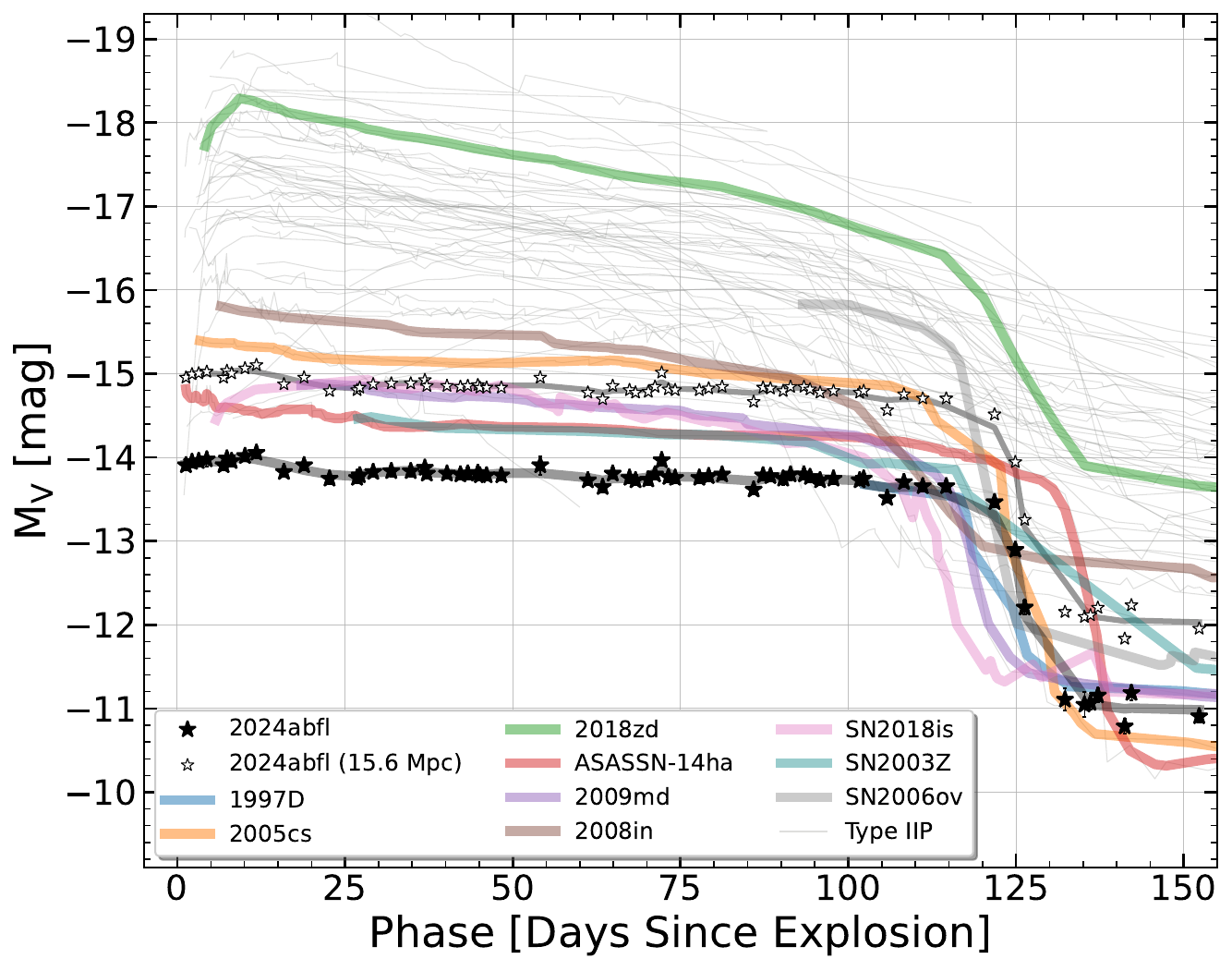}}
       \caption{$V-$band light curve evolution for SN~2024abfl compared with other low-luminosity SNe from the literature. Grey lines in the background show a larger Type II SNe sample. We also plot $\rm M_V$ evolution of SN~2024abfl for the higher distance (15.6~Mpc).}
    \label{fig:lccomp} 
   \end{figure}

We present the light curve evolution sequence for all the UV and optical filters in Figure~\ref{fig:lightcurves}. The evolution spans up to $\sim$ 180~d after the explosion. Optical light curves are followed immediately after discovery, whereas UV light curves begin several days later. We do not find a clear rising trend in the light curve evolution, both in UV and in optical bands, suggesting a very steep rise to the peak within a few days after the explosion. Considering this, the true peak magnitude or rise time could not be determined for SN~2024abfl. Further evolution reveals very flat light curves, and the apparent decline in the light curves is almost non-existent in redder bands. In Type IIP supernovae, a gradual transition from the plateau to the radioactive tail phase is observed, with a typical duration of $\rm \sim 20~days$ \citep{2018avinash}. However, in SN~2024abfl, the transition is achieved relatively sooner ($\rm \sim 10~days$). We estimate the plateau duration of SN~2024abfl to be $126\pm1$~d post explosion. The plateau length was estimated using the $\rm t_{PT}$ definition from \citet{2016MNRAS.459.3939Valenti} and by fitting their equation with \textit{emcee}.

Figure~\ref{fig:lccomp} shows the $V$-band light curve comparison of SN~2024abfl with other Type II SNe. We also compare SN~2024abfl with other well-studied intermediate to low-luminosity Type II SNe: SN~1997D \citep{2000A&A...354..557C, 2003MNRAS.338..711Z}, SN~2005cs \citep{Pastorello2006_2005cs, Faran2014_2005cs}, ASASSN-14ha \citep{2016MNRAS.459.3939Valenti}, SN~2009md \citep{2011MNRAS.417.1417F}, SN~2008in \citep{2008inBose}, SN~2003Z \citep{2014LLIIP}, SN~2006ov \citep{2014LLIIP}, and SN~2018is \citep{2025A&A...694A.260D}, and find SN~2024abfl to have the lowest inferred brightness. The mid-plateau absolute  $V$-band magnitude is found to be $\rm-13.8\pm0.10$~mag. Only SN with lower magnitude in literature is SN~1999br \citep{2004MNRAS.347...74P} with  $\rm-13.7$~mag. The distance modulus used for SN~1999br was 31.19~mag, whereas recent redshift-independent distance estimates are slightly higher, ranging from 31.60 to 32.98~mag \citep{2010ApJ...715..833O}, with 32.11~mag being the most precise \citep{2014AJ....148..107R}. Considering these distance estimates, the $V$-band magnitude of SN~1999br becomes brighter by 0.4-1.7~mag, making SN~1999br brighter than SN~2024abfl. Hence, it leaves SN~2024abfl as one of the faintest Type IIP SN discovered to date. \citet{2026arXiv260102638G} adopted a distance of 15.6 Mpc for NGC~2146. Our EPM analysis and a previous independent SCM analysis from SN~2018zd suggest a significantly lower value, implying a substantially lower intrinsic brightness for SN~2024abfl. Even for the higher distance estimate, the nature of SN~2024abfl remains sub-luminous, albeit the SN itself does not remain the faintest. 

Apart from the faint nature of SN~2024abfl, its plateau decline is also intriguing, where the decline is almost non-existent. We estimate a very low decline rate of $0.10\pm0.04$ mag/100~days, which again is one of the slowest among all well-studied Type II SNe. The slowest decline rate we could estimate in the sample is observed in SN~2005cs ($0.04\pm0.02$ mag/100~days), but this decline rate is based on only a fraction of the full light-curve period and is not averaged over the full plateau duration. The light-curve decline is very low, but this is not unique to SN~2024abfl, as other low-luminosity Type II SNe also exhibit a very slow-declining plateau phase. For example, ASAS-SN~14ha ( $0.34\pm0.09$ mag/100~days), SN~2009md ( $0.57\pm0.17$ mag/100~days), SN~2018is ( $0.58\pm0.16$ mag/100~days), show a decline rate around $0.50$~mag/100~days or less. Other SNe in the sample decline faster than those mentioned above. In many cases, especially for Type IIP SNe, brighter SNe tend to decline faster and have higher explosion energies \citep{2014Anderson}.  
\begin{figure}[hbt!]
     \resizebox{\hsize}{!}{\includegraphics{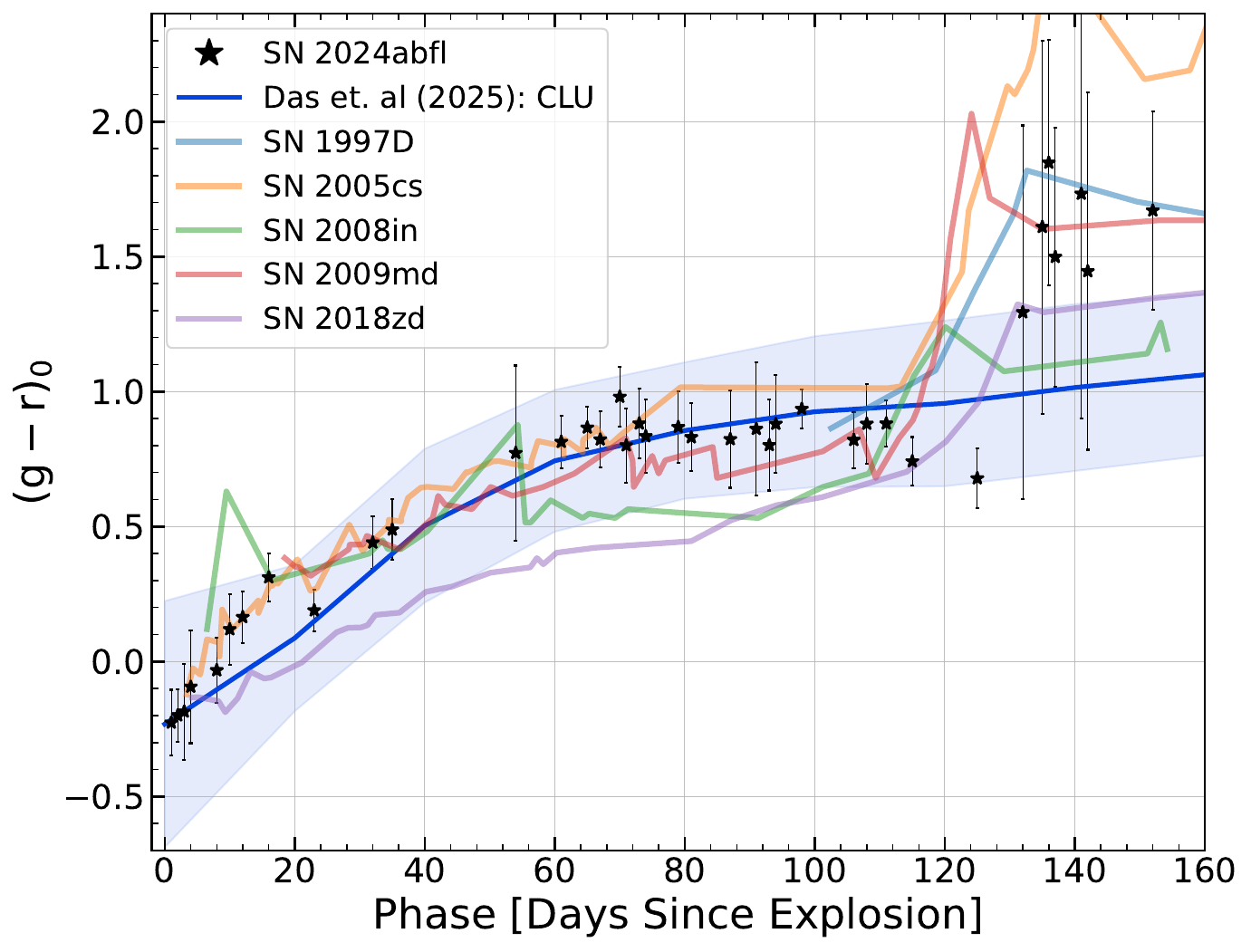}} \\
     \centering
      \resizebox{0.95\hsize}{!}{\includegraphics{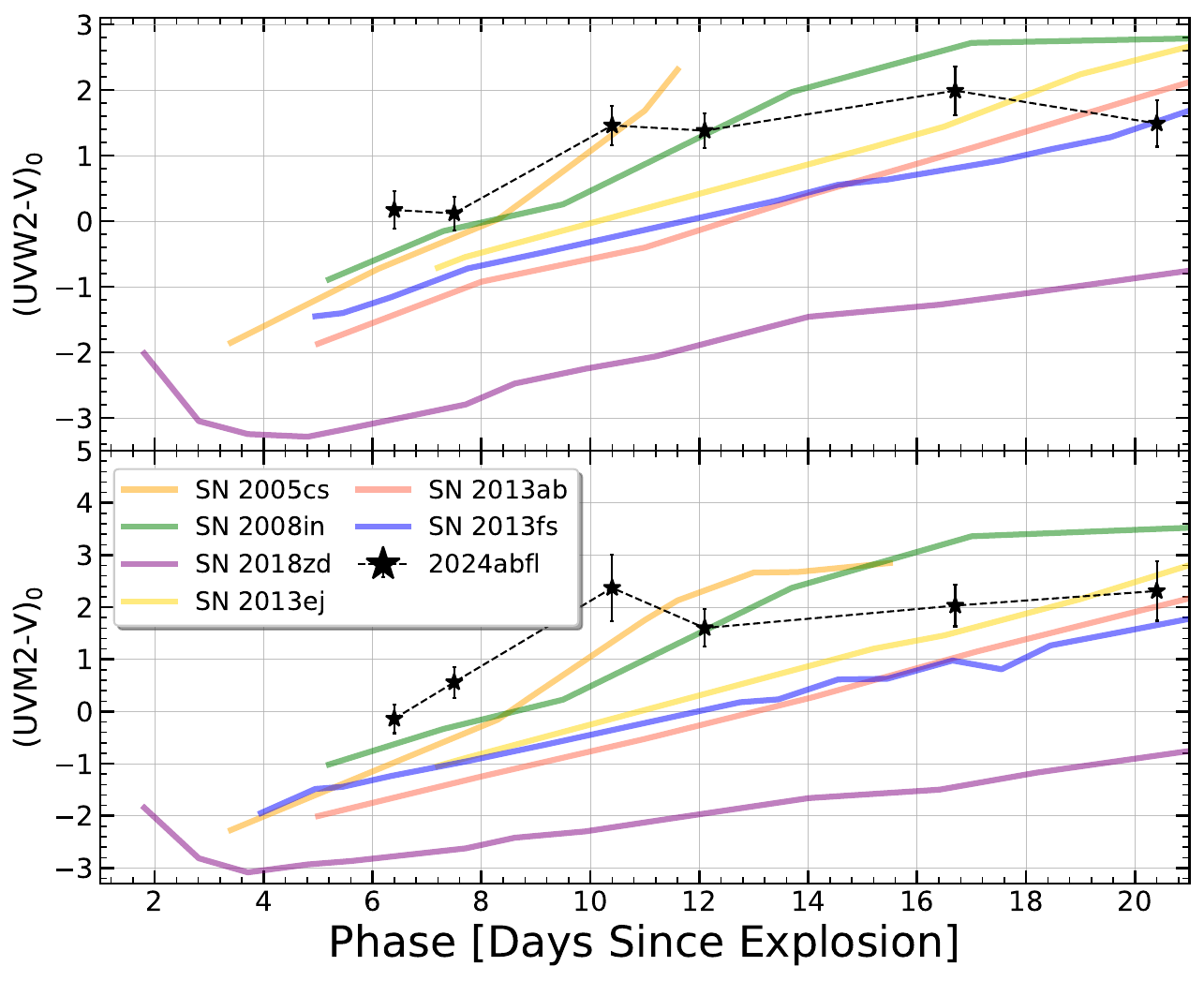}}
         \caption{ \textbf{Top:} $g-r$ color evolution of SN~2025abfl compared with other well known low-luminosity SNe. We also plot the mean and 1-sigma values obtained for a large sample of Type II SNe (including LLIIP) present in the census of the local universe (CLU) sample \citep{2026PASP..138b4204D}.
         \textbf{Bottom:} UV-Optical colors for SN~2024abfl along with some typical SNe including low-luminosity SNe are shown.} 
    \label{fig:colors} 
     \end{figure}

In the upper panel of Figure~\ref{fig:colors}, we plot the $(g-r)_0$ color evolution of SN~2024abfl. The $(g-r)_0$ color trend is blue for initial $\sim$~10~d and thereafter turns redder. We compare the color with the mean color obtained from a larger Type II SNe sample \citep{2026PASP..138b4204D}. The $(g-r)_0$ color evolution is fairly normal till the end of the plateau and follows the mean trend line within 1-$\sigma$ as obtained for the larger sample of Type II SNe, including the low-luminosity ones. As the plateau reaches its end, we find $(g-r)_0$ color becomes slightly bluer and then becomes red abruptly. As seen in the Figure, this abrupt transition to a redder color is not unique to SN~2024abfl. We observe this change in color for other low-luminosity Type II SNe as well.  This trend is missing from the larger Type II sample, probably due to the fact that the larger sample covers different plateau lengths, and the effect is diminished while averaging. This `elbow' like feature also appears for a small sample of Type IIP SNe as observed in $V-R$ colors from \citet{2023ApJ...954..155T}, however, with a gradual transition of $\rm \sim 20~days$. 

In the bottom panel of the same Figure~\ref{fig:colors}, we plot UV-optical color evolution for SN~2024abfl. We find that the color evolution is similar to that of low- to intermediate-luminosity objects such as SN~2005cs \citep{Faran2014_2005cs} and SN~2008in \citep{2008inBose, 2024IAUS..361..610T}. SNe showing interactions are much bluer than SN~2024abfl. However, with limited early UV data, we cannot rule out early CSM interaction \citep{2026arXiv260102638G, 2026arXiv260204309C}.

\begin{figure}[htb!]
    \centering
    \resizebox{\hsize}{!}{\includegraphics{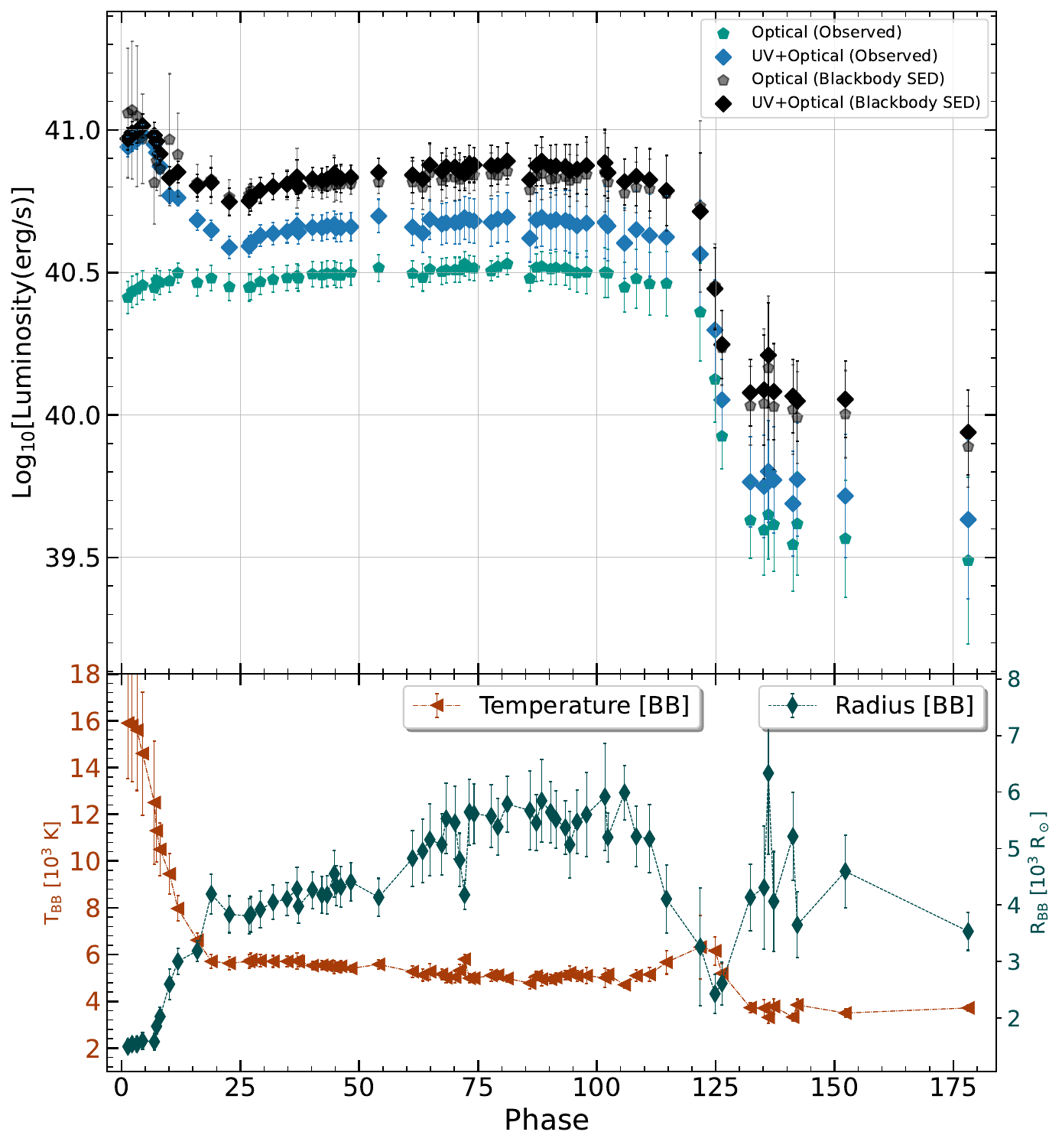}}
    \caption{Top: Pseudo-bolometric/bolometric light curve evolution for different filter sets as obtained using Superbol. Bottom: Blackbody radius and temperature evolution obtained by fitting the photometry flux with a blackbody SED. }
    \label{fig:bolom}
\end{figure} 

To estimate the $\rm ^{56}Ni$ mass, we used multi-broadband photometry and created a pseudo-bolometric light curve of SN~2024abfl. For this purpose, we used the \texttt{SuperBol} code. The code computes pseudo-bolometric/bolometric curves by integrating the flux over observed bands. We carried it out in two ways, including the UV filter data and without UV data. We further note that the luminosity values obtained from a blackbody SED applied to the optical and UV+Optical data separately are very similar (Figure~\ref{fig:bolom}). This could imply that the amount of CSM presence is very low and inconsequential to the blue excess (see Blackbody fits to the optical data in Appendix~\ref{appendix:1}). We further note that with no near-UV coverage during the first week post-explosion, the possibility of flux enhancement due to CSM could not be completely ruled out. We regard the UV+Optical light curve as the bolometric light curve of SN~2024abfl in further analysis.  Further, we note that within the first three weeks, the black-body temperature drops to 5000~K, starting from around 17000~K, around the time of discovery. This temperature is on the lower end for typical Type II SNe, but is considered atypical in low-luminosity Type II SNe \citep{Faran2014_2005cs, 2025A&A...694A.260D}.

The $\rm ^{56}Ni$ decay chain primarily dominates the late-time light curve evolution of Type II SNe. It is the primary energy source during the nebular phase of Type II SNe. We used various methods to estimate the mass of synthesized $\rm ^{56}Ni$. We compared the bolometric luminosity of SN~2024abfl in the nebular phase with that of SN~1987A \citep{ArnettFu1989}. The mass of $\rm ^{56}Ni$ in  SN~1987A is very well constrained using multiband photometry and hydrodynamical modeling and can be utilized to estimate the mass of $\rm ^{56}Ni$ in SN~2024abfl. We compare the bolometric luminosity with the values obtained for SN~1987A at similar epochs and use a simple scaling (Equation~\ref{eq:Nickel1987A}) to get an estimate on $\rm ^{56}Ni$ mass.

\begin{equation}
\label{eq:Nickel1987A}
    M_{Ni}(SN) \approx M_{Ni}(SN~1987A)\cdot \frac{L_{SN}(t)}{L_{1987A}(t)}
\end{equation}
From the late time light curve ($\rm >130~d$), we compared at 4 similar epochs and estimated the mass of $\rm ^{56}Ni$ to be $\rm M_{Ni} = 0.0039\pm0.0006~M_\odot$. Another way to estimate $\rm ^{56}Ni$ mass is by utilizing radioactive decay properties, as given in Equation~\ref{eq:Nitc}.
\begin{equation}
    \label{eq:Nitc}
    L_{obs}(t) = L_0M_{Ni}\left [ e^{-\frac{t}{t_{Co}}}-e^{-\frac{t}{t_{Ni}}}\right ]\left ( 1-e^{-\frac{t_c^2}{t^2}}\right),
\end{equation} where the constants $L_0$ is $\rm 1.41\times10^{43} erg~s^{-1}$, $t_{Co}$ (111.4~d), and $t_{Ni}$ (8.8~d) are the respective characteristic decay times. Here,
$t_c$ is the time when the optical depth for
$\rm \gamma-$rays approaches unity \citep{2016MNRAS.461.2003Y}. We obtained  $\rm ^{56}Ni$ mass and characteristic time of $\rm 0.0030\pm0.0001~M_\odot$ and $>655\pm80~d$, respectively. The characteristic timescale ($t_c$) was poorly constrained, indicating little to no $\rm \gamma-$ray leakage. $\rm ^{56}Ni$ mass estimates from both the methods corroborate each other, with $\rm M_{Ni}\approx0.0035(6)~M_\odot$ being the average value. However, if we consider the light curve obtained using the larger distance (15.6~Mpc), $\rm ^{56}Ni$ mass could go as high as 0.008~$\rm M_\odot$. \citet{2026arXiv260102638G} and \citet{2026arXiv260204309C} also obtained higher $\rm ^{56}Ni$ mass as per the assumed larger distance. 

\begin{figure}[htb!]
    \centering
        
    \resizebox{\hsize}{!}{\includegraphics{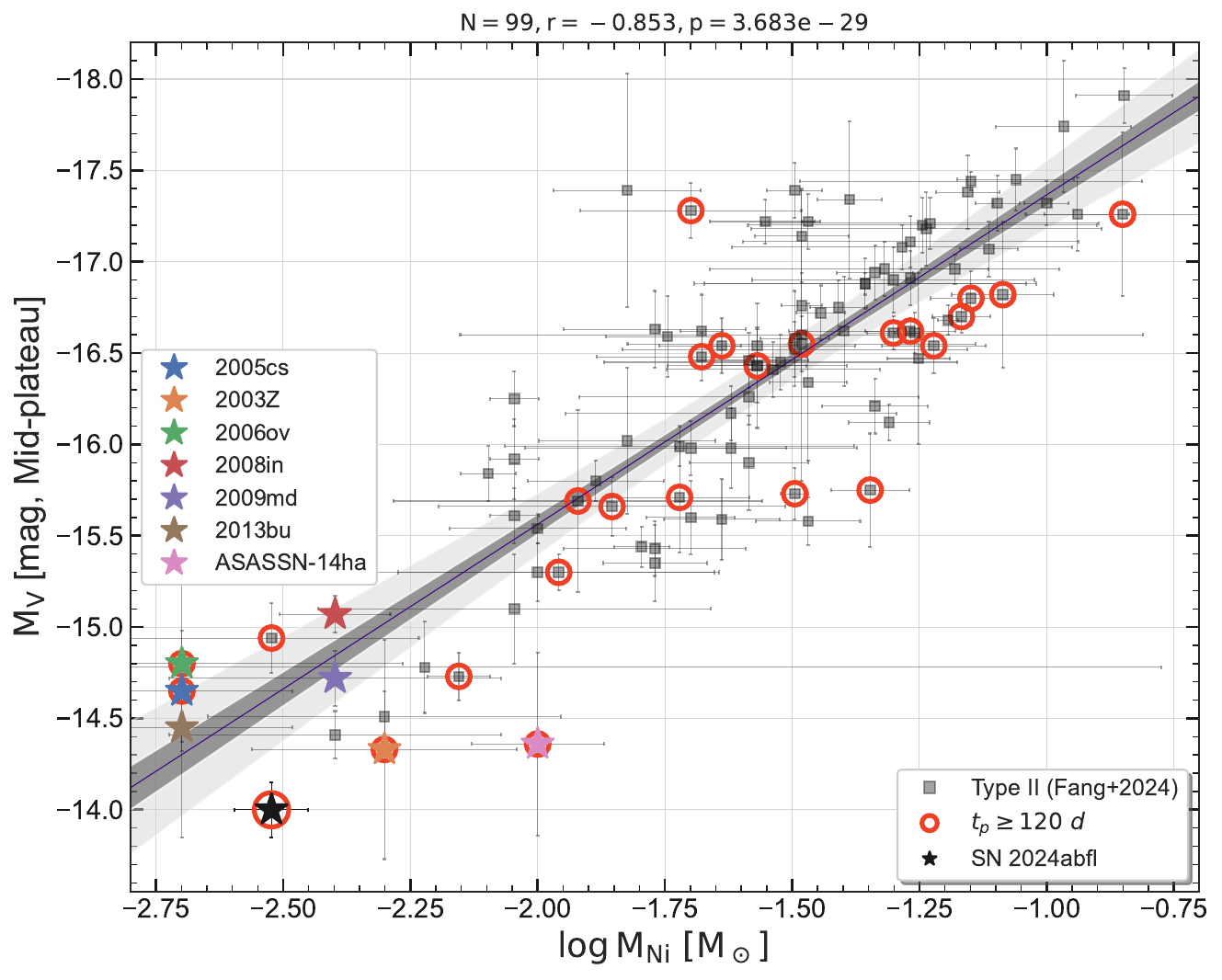}}

    \caption{Nickel mass, $\rm M_{Ni}$ vs mid-plateau brightness, $\rm M_V$ plot for a sample of Type II SNe obtained from \citet{2024arXiv240401776F}. Some of the low-luminosity events are also marked separately. Additionally, SNe with longer plateau durations ($\rm t_p\geq120~d$) are encircled in red color.}
    \label{fig:corr}
\end{figure}

In Figure~\ref{fig:corr}, we show SN~2024abfl position over the well-established luminosity (around mid-plateau) vs nickel mass correlation for a large Type II sample \citep{2014Anderson, 2024arXiv240401776F}. An optimal linear fit is also shown with 1- and 2-$\sigma$ scatter. Along with this, we try to differentiate the SN events which have longer plateau durations ($\rm t_p\geq 120~d$). Clearly, SN~2024abfl lies at the extremes of this sample. It is noteworthy that a significant number of low-luminosity SNe do not have extended plateau lengths; equally, a good number of longer plateau duration events are brighter with higher mass of estimated $\rm ^{56}Ni$ masses.   

\begin{figure*}[htb!]
    \centering
        
    \resizebox{0.85\hsize}{!}{\includegraphics{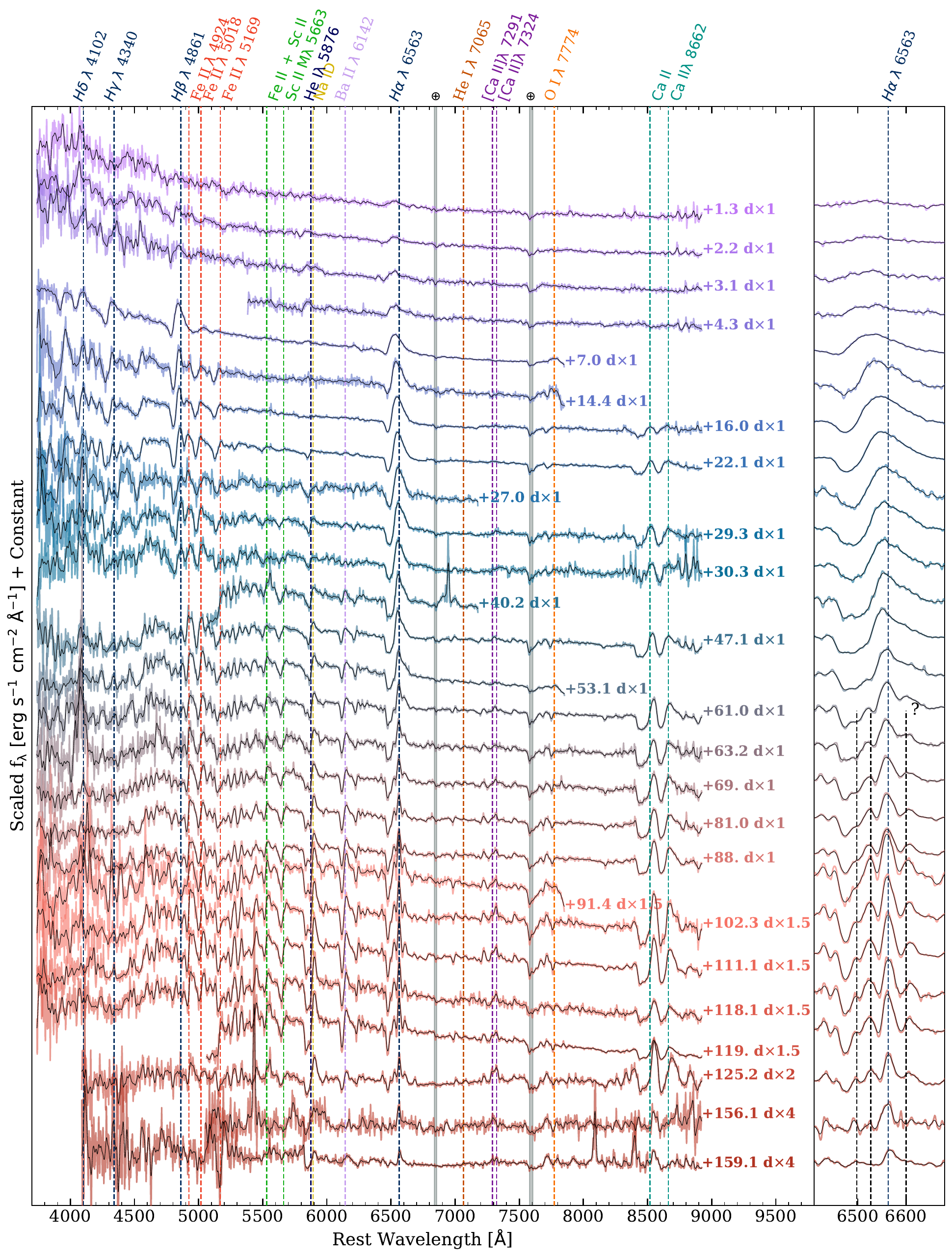}}
        \caption{The entire spectral evolution for SN~2024abfl corrected for redshift, extinction, and scaled with respect to photometric flux is shown here. On the right side, we show the evolution of H$\alpha$. (Constant offsets have been applied to the individual spectra for visual clarity. Well-identified lines have been marked. The number next to the phase represents the scaling coefficients applied to the spectra.) }
    \label{fig:spec}
\end{figure*}

\section{Spectra}
\label{sec:spec}

The entire spectroscopic sequence corrected for redshift, multiband photometry, and extinction is presented in Figure~\ref{fig:spec}. Some of the well-known Type II SNe lines, which are clearly identified, have also been marked in the same figure. The spectra cover 160~d of SN evolution. From the very first spectrum at $+~$1.3~d, we find the emergence of several emission lines with the clear identification of hydrogen Balmer features ($\lambda \lambda \lambda \lambda~ 6563,~4861,~4340,~4102$). These lines become much more conspicuous around $+~$7.0~d, and their profile shows P-Cygni behavior, typical of Type II SNe. Around the same phase, the bluer side of the spectrum begins to be dominated by metal features, especially those from Fe lines.  At $+~$14.4$-$16.0~d, we could clearly observe Fe lines $\lambda\lambda\lambda~4924,~5018,~5169$ and Ca NIR features. As the SED becomes much cooler, other metal lines also start to appear. Around the mid-plateau phase, the spectra are dominated by metal lines from Ba, Fe, Sc, and Na. Starting at the same phase ($+~$61~d), we find peculiar behavior in the H$\rm \alpha$ profile where we observe two separate peaks emerge on either side of its central peak. Both components are shifted by $\rm \approx\pm 1800~km~s^{-1}(40~$\AA). In literature, these are associated with either Ti~II \citep{2004MNRAS.347...74P} or Ba~II  $\rm\lambda~6497$ \citep[SN~1997D,][]{2004MNRAS.347...74P}. However, we could not find any Ti II lines in this region in the comprehensive line list of elements \citep{NIST_ASD}. Similarly, for Ba~II, these lines appear at different wavelengths, particularly at $\rm \lambda\lambda~6526,~6600$, which are more plausible for Ba~I $\rm \lambda\lambda~6527,~6595$ within our wavelength resolution of $\sim$10~\AA. Such features on either side of $\rm H\alpha$ could also be the result of emission originating from a cold dense shell \citep{2016MNRAS.457.3241A, 2026arXiv260314137S}. Usually, these features implying dust formation appear during the late nebular phases of typical Type IIP SNe with early CSM interaction and are accompanied by other changes in the line profiles \citep{2025A&A...698A.293D}. These features are not observed in the post-plateau phase of SN~2024abfl. However, early CSM signatures \citep{2026arXiv260102638G} leading to these features at this stage could not be completely ruled out. Detailed spectral modeling could discern these features with greater confidence, which is beyond the scope of this work.      

\begin{figure}[htb!]
    \centering
    \resizebox{\hsize}{!}{\includegraphics{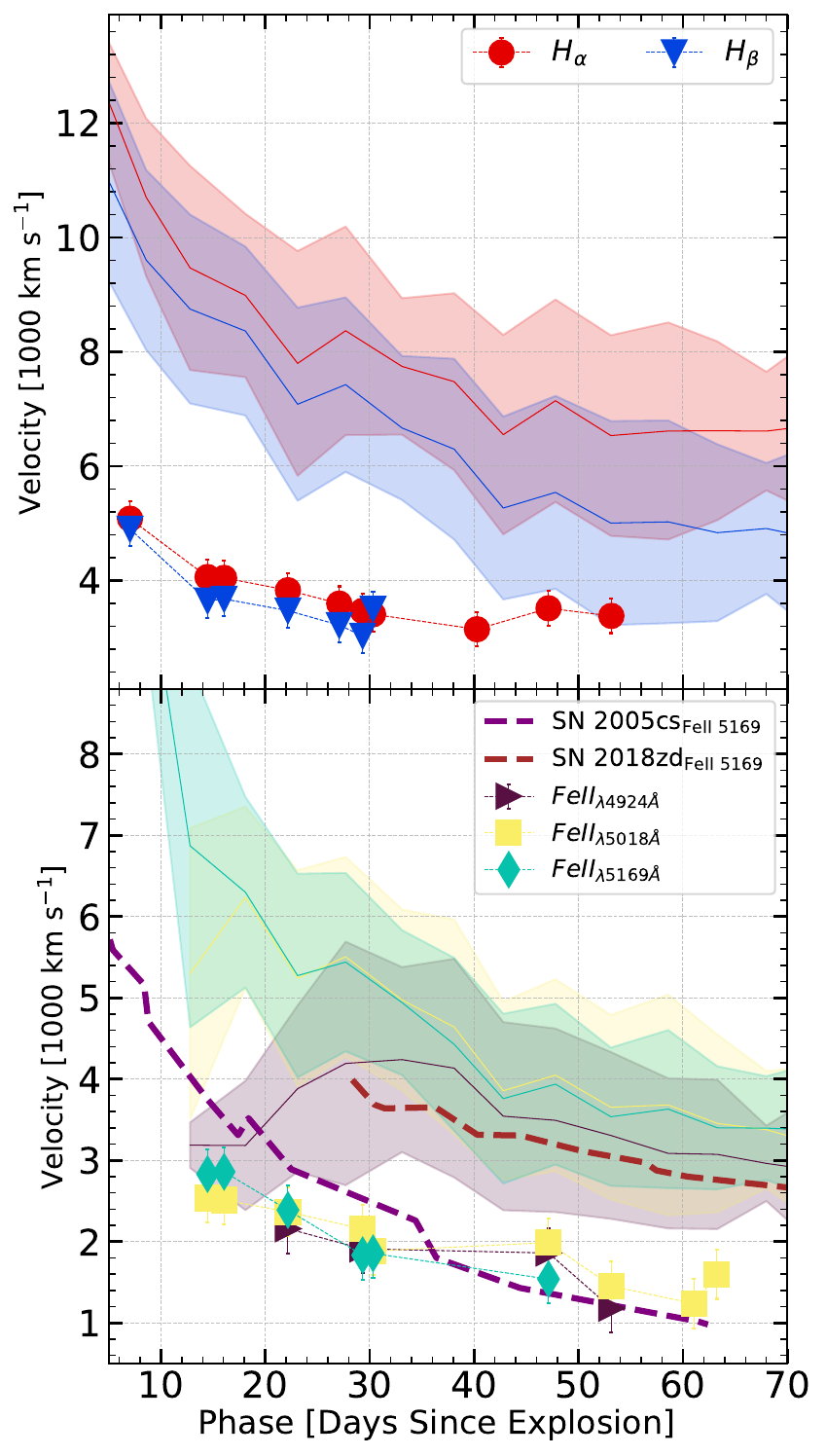}}
    \caption{Velocity evolution obtained for several prominent and isolated metallic features (including Balmer features) observed in the spectra is shown here. The continuous curves show mean velocities for a large Type II sample \citep{Gutierrez2017_TypeIISample}
with the corresponding $\rm 1-\sigma$ scatter shaded around the mean. }
    \label{fig:vel}
\end{figure}
We estimate the expansion velocities using several prominent absorption features and plot these in Figure~\ref{fig:vel}. We find that the expansion velocities are very low and evolve slowly throughout. The slow evolution and low expansion velocities become more obvious when compared to a large Type II sample obtained from \citet{Gutierrez2017_TypeIISample}. The evolution of SN~2024abfl velocities is significantly lower than the 1-$\sigma$ scatter for a larger sample of Type II SNe. After $+$20~d, the velocity evolution becomes flatter and settles around 1500-2000~$\rm km~s^{-1}$ for Fe II triplet.     

 \begin{figure*}[htb!]
        \centering
    \resizebox{\hsize}{!}{\includegraphics{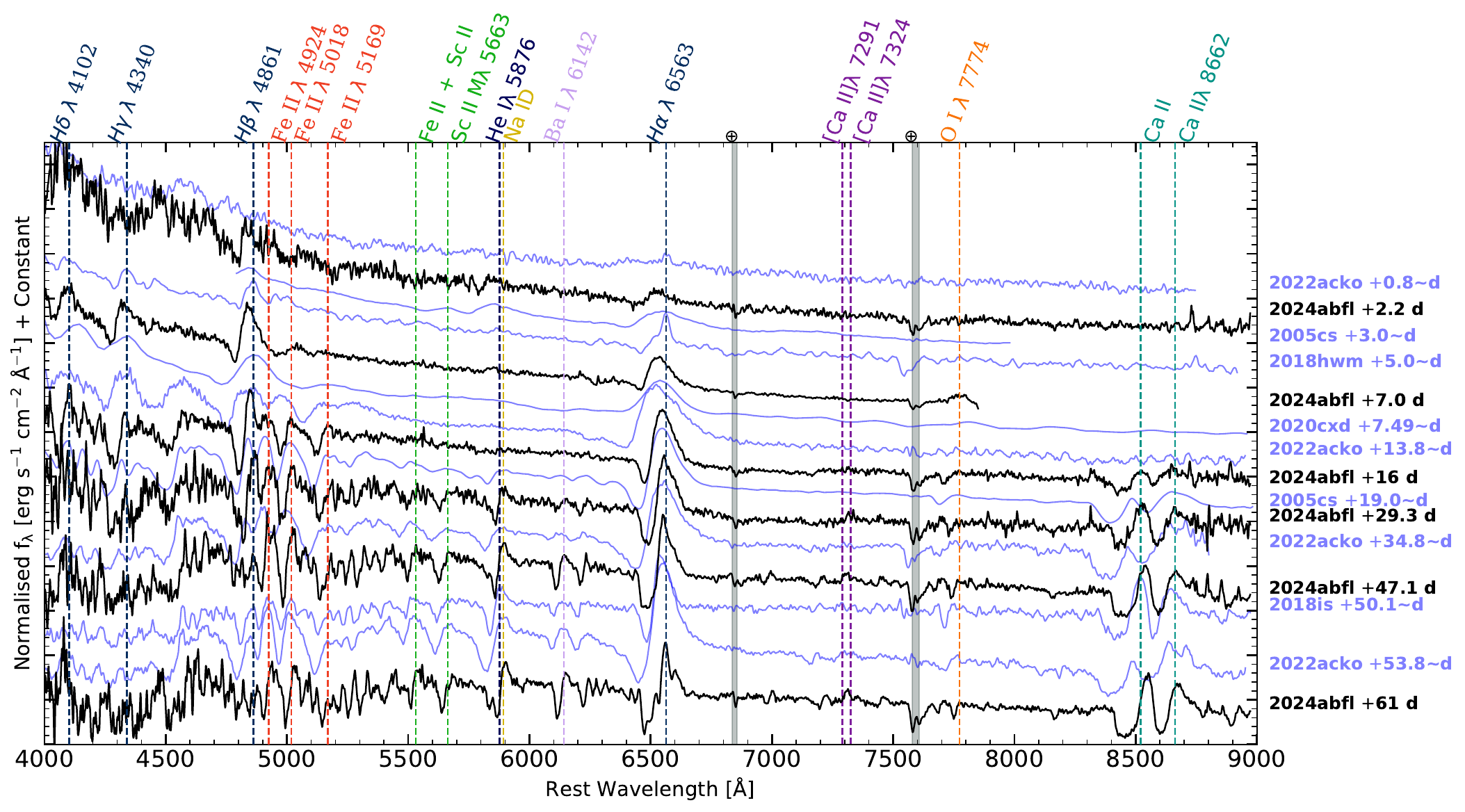}}\\
    \caption{Comparison of SN~2024abfl spectra with the spectra of other well-studied low-luminosity Type II SNe sequentially covering up to the mid-plateau phase. }
    \label{fig:speccomp1}
\end{figure*}

\begin{figure*}[htb!]
    \resizebox{\hsize}{!}{\includegraphics{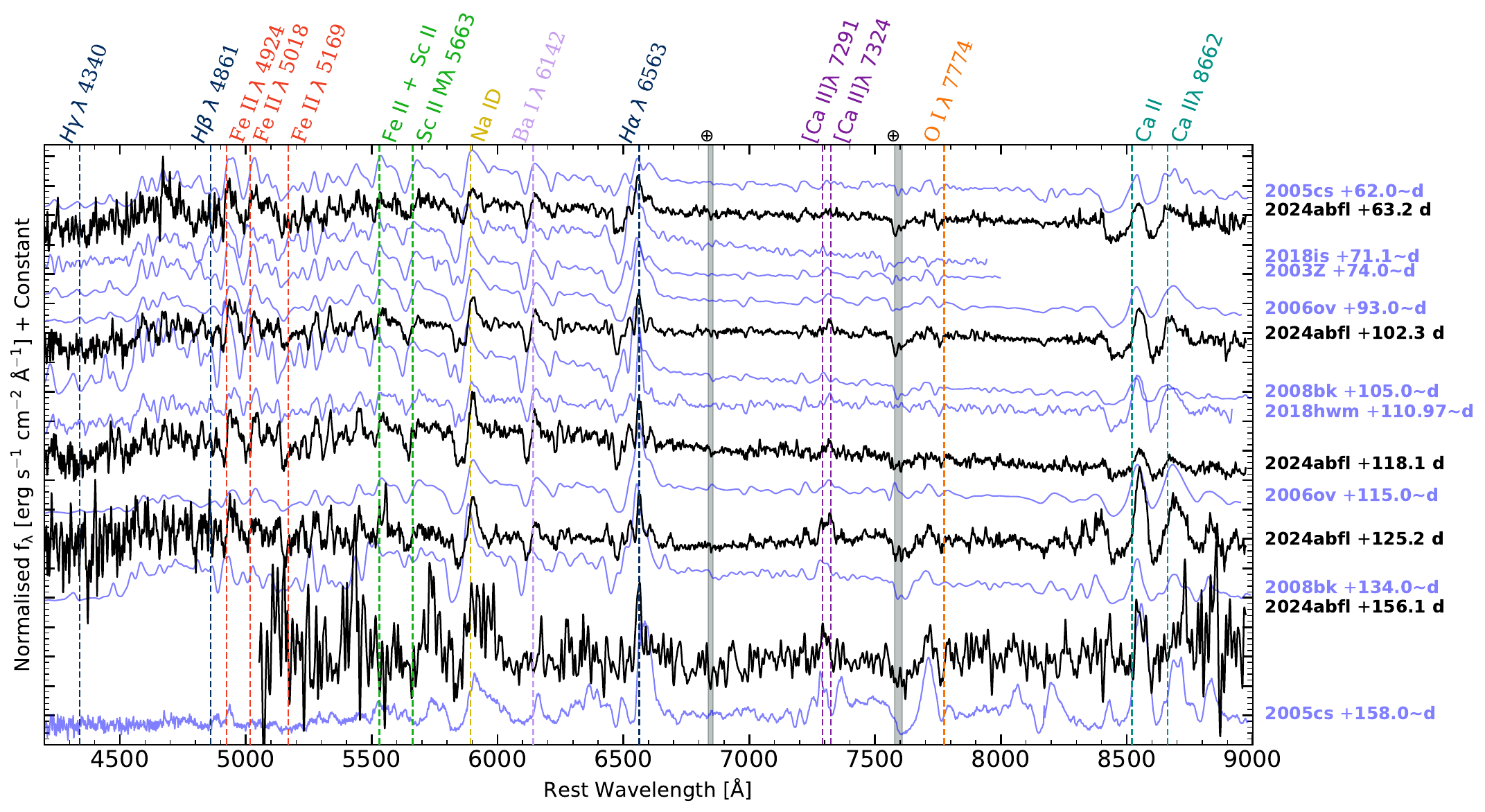}}

    \caption{Comparison of SN~2024abfl spectra with the spectra of other well-studied low-luminosity Type II SNe sequentially from mid-plateau to the radioactive tail phase.}
    \label{fig:speccomp2}
\end{figure*}

In Figure~\ref {fig:speccomp1}, we compare the spectral evolution of SN~2024abfl sequentially with some of the well-studied, low-luminosity Type II SNe. These low-luminosity SNe cover some of the earlier as well as recent objects, including the archetypal SN 2005cs \citep{Pastorello2006_2005cs, 20062005cs, 20092005cs, Faran2014_2005cs}. The comprehensive list includes: SN~2003Z \citep{2014LLIIP}, SN~2006ov \citep{2014LLIIP}, SN~2008bk \citep{2023A&A...678A..43N}, SN~2018is \citep{2025A&A...694A.260D}, SN~2018hwm \citep{2021MNRAS.501.1059R}, SN~2020cxd \citep{20212020cxd, 2022MNRAS.513.4983V}, and SN~2022acko \citep{2023ApJ...953L..18B}. We obtained spectral time-series data for these objects from the Wiserep\footnote{\url{https://wiserep.org}} archives \citep{2012PASP..124..668Y}. Broadly, across all phases, we find the spectral evolution to be uniquely rich, with a plethora of metallic features, along with similarities and differences with the compared SNe. The earliest spectrum from the comparison sample is available for SN~2022acko, taken approximately $+~$1~d after the explosion, and is predominantly a featureless continuum, with hints of the development of several features. The spectrum of SN~2024abfl around $+~$2~d is slightly bluer and feature-rich with conspicuous Balmer lines. In the early phase, SN~2018hwm shows an additional narrow component over broad H$\alpha$ emission. SN~2005cs, around a similar phase, shows broader Balmer features. Even after 16 days, the spectral features of SN~2024abfl are significantly narrower with deeper absorptions compared to the other SNe, signifying low underlying expansion velocities. Around $+~$20 days, the spectral signatures and continuum shape replicate those of SN~2022acko and SN~2005cs, but with much narrower absorption and emission components in SN~2024abfl. Until mid-plateau ($\sim$~60~d), SN~2024abfl's spectral behavior is very similar to that of other low-luminosity counterparts, such as SN~2022acko, SN~2018is, and SN~2005cs. 

In Figure~\ref{fig:speccomp2}, we continue the spectral comparison with the aforementioned sample evolving from mid-plateau onward to the nickel-decay powered phase. We note that the emergence of two emission peaks around the H$\alpha$ rest wavelength is a common phenomenon in low-luminosity SNe, which typically becomes conspicuous during the mid-plateau phase. The only exception we found in the sample is SN 2006ov, which shows hints of emission lines on either side of H$\alpha$, though not as prominent as in other low-luminosity SNe and SN~2024abfl. These emission features fade after the plateau phase is over and are not detected in other supernovae either. Whether these features are a result of either lower ejecta temperatures or lower expansion velocities, or a combination of both, is subject to detailed spectral modeling. Nevertheless, the spectra of low-luminosity Type II SNe are remarkably rich in various metallic features, both from neutral and ionized species. During the late plateau period, the blue region of the spectrum is so dominated by metallic features that \ion{Fe}{2} lines almost blend with them.    

\section{Progenitor}
\label{sec:origins}
\begin{figure}[htb!]
    \centering
    \resizebox{\hsize}{!}{\includegraphics{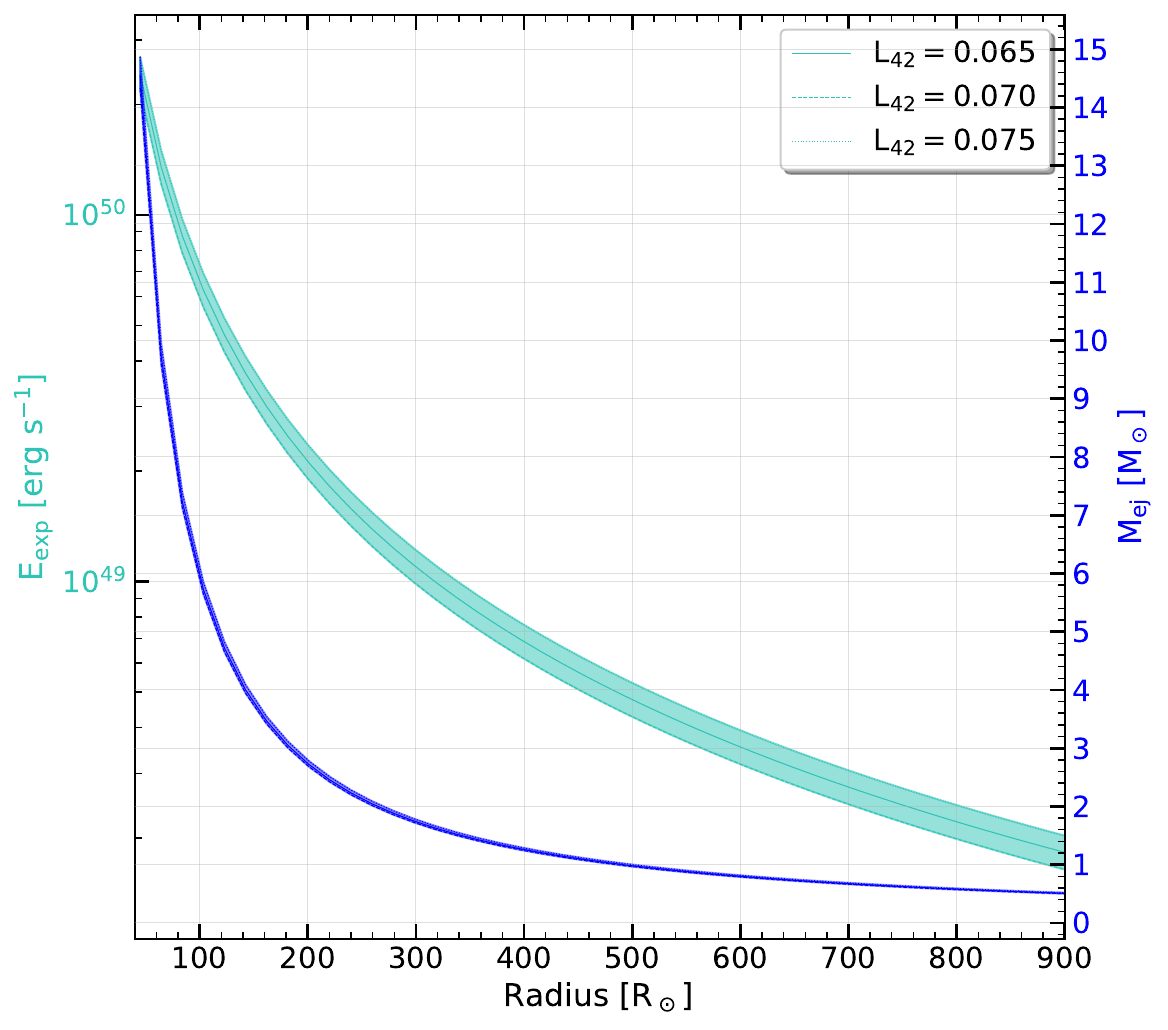}}
    \caption{ Expected ranges of  $E_{exp}$ and $M_{ej}$ obtained from the scaling relations derived in \citet{2019ApJ...879....3GGOLD}. The shaded regions include the values obtained considering the errors in the observables.}
    \label{fig:mesa-scaling}
\end{figure}

It has been a longstanding challenge to zero-in on the progenitor of low-luminosity SN events. Several objects studied in the past or in recent studies have also yielded a wide range of progenitor masses. Previous studies have shown that low-luminosity SNe can arise from a range of progenitor masses, from low- to moderate- to high-mass RSGs \citep{2014LLIIP, 2017MNRAS.464.3013P, 2021MNRAS.503..797K,2022ApJ...934...67B}. With high mass RSGs showing significant fallback, as also observed in SN~20005cs \citep{Paxton2018} and SN~2021wvw \citep{2024ApJ...974...44T}. Prior constraints on the progenitor mass of SN~2024abfl come from a recent study by \citet{2025ApJ...982L..55L}, which utilized archival HST images. Based on their analysis, SN~2024abfl could arise from a plausible RSG with a mass of 10-16~$\rm M_\odot$. We attempt to explore progenitors within this mass range, further narrowing down the probable progenitor's mass range and estimating other explosion properties. 

\begin{table}[hbt!]

\caption{Key properties for several of our pre-SN RSGs evolved using MESA}
\centering
\begin{tabular}{ccccc} \hline
$\rm M_{ZAMS}$          & $\alpha_{MLT}$         &$\rm M_{Fe-Core}$    & Radius & $\rm T_{eff}$        \\        
      ($M_\odot$) & & ($M_\odot$) & ($R_\odot$) & ($K$) \\
\hline
15 & 1.5 &  1.81& 1386& 3106 \\
12 & 3.0 & 1.49 & 668& 3892 \\
10 & 1.5& 1.51 & 883& 3122  \\
10 & 2.0& 1.51 & 716& 3424 \\
10 & 3.0& 1.54 & 536& 3960  \\
9 &3.0 & 1.55 & 466 & 3960 \\
9 & 4.0& 1.56 & 374& 4417 \\
        
\hline
\end{tabular} \\
\label{tab:mesa_param}
\end{table}

\begin{figure*}[htb!]
    \centering
        
    \resizebox{\hsize}{!}{\includegraphics{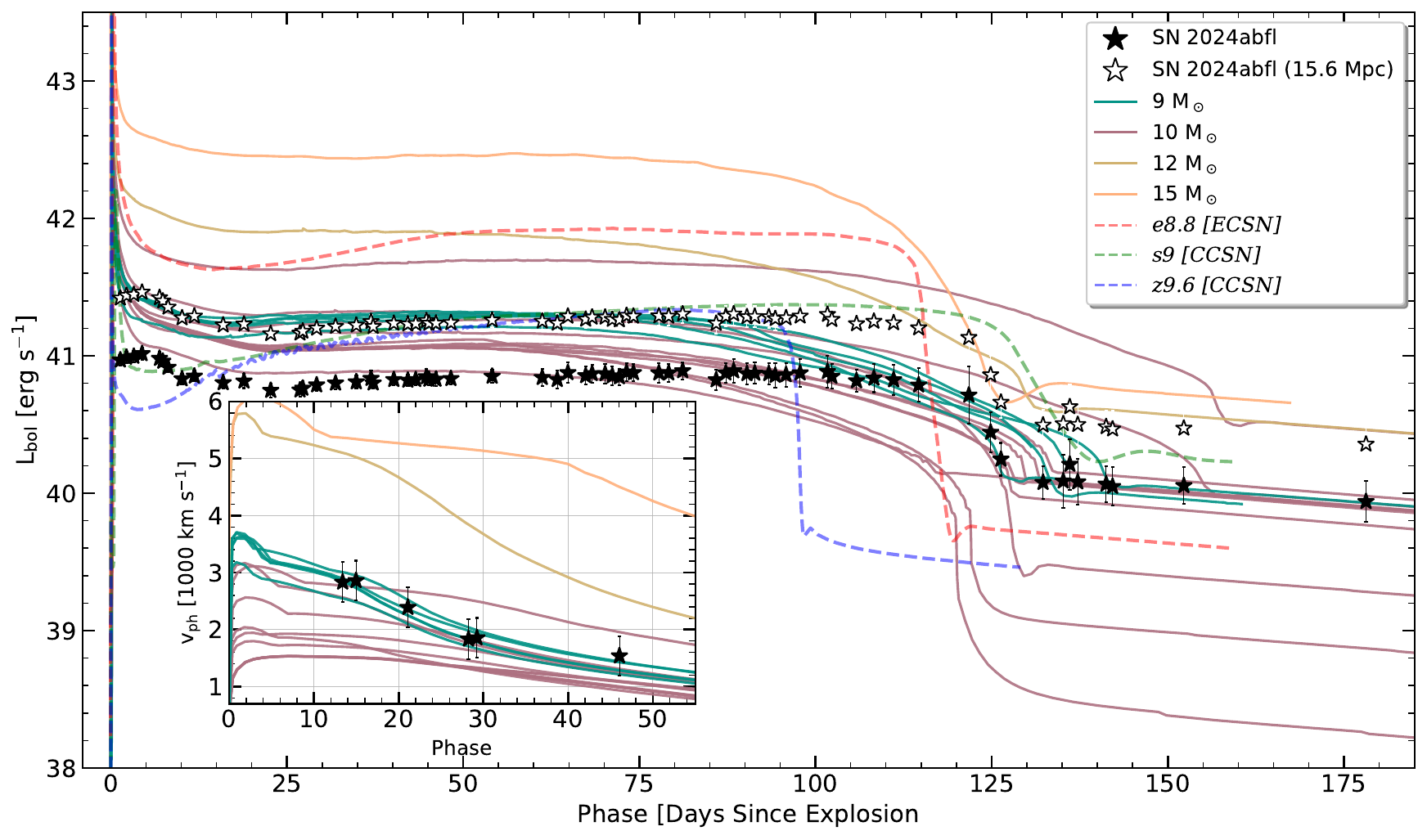}}

    \caption{Model light curves for 9-15 $\rm M_\odot$ ZAMS masses obtained using MESA+STELLA framework are compared with the observed bolometric luminosity (UV+Optical) of SN~2024abfl. Various 9 and 10~$\rm M_\odot$ models also differ by explosion energy (0.01-0.1 foe) and $\rm ^{56}Ni$ mass (0.003-0.008~$\rm M_\odot$). The inset shows the corresponding photospheric velocities obtained for various progenitor models. ECSN and low-energy Fe-CCSN model light curves from \citet{2021MNRAS.503..797K} are also overplotted.}
    \label{fig:mesa}
\end{figure*}

For the initial explosion energy and ejecta mass ranges, we use scaling relations obtained by  \citet{2019ApJ...879....3GGOLD}. We obtained these expected quantities for radii ranging from 50~$\rm R_\odot$ to 900~$\rm R_\odot$, which are shown in Figure~\ref{fig:mesa-scaling}. These ranges are valid for some very compact progenitor cases to typical RSG progenitors.  These relations depend on the plateau luminosity, plateau length, synthesized nickel mass, and the radius of the progenitor. SN~2024abfl, being a low luminosity SNe with a meager amount of synthesized nickel mass, has explosion energy and ejecta mass on the lower end. We obtained $\rm E_{exp} \leq 10^{49}~erg$ and $\rm M_{ej} \leq 2~M_\odot$, for typical RSG radii greater than 300~$\rm R_\odot$. Evidently, we obtain degenerate solutions from these relations. However, several outcomes are noteworthy or provide a good indication for the plausible progenitor properties: One is the very low explosion (order $10^{49}-10^{50}$~erg) energies, and the other factor is the compactness of the progenitor for a viable solution. Because the typical or extended progenitors lead to very low ejecta masses ($\leq~2~M_\odot$), whereas the plateau length obtained for SN~2024abfl is significantly longer and is not reproducible by such low ejecta masses. 

It is challenging to explode massive progenitors with very low energies, leading to significant fallback of the material onto the core, as it does not overcome the gravitational potential energy required to fully disrupt the ejecta. It had been observed in more sophisticated simulations that massive progenitors are usually associated with higher energies, but explosion energy is not a monotonic function of mass \citep{2021Natur.589...29B}. We also explored models from different light curve grids to determine whether they could reproduce the observed behavior of SN~202abfl. We checked light-curve matches from various published datasets \citep{Morozova2015, 2023PASJ...75..634M}, which also included low explosion energies and include low-mass progenitors. However, these models were unsatisfactory in reproducing the observed behavior of SN~2024abfl. Hence, we evolve several new progenitor models to better constrain the explosion and progenitor properties for SN~2024abfl. For this purpose, we made use of \texttt{MESA} \citep{Paxton2011,Paxton2013,Paxton2015,Paxton2018,Paxton2019,Jermyn2023} revision 24.08.1 to evolve the progenitor and hydrodynamical explosion and \texttt{STELLA} \citep{Blinnikov2004,Baklanov2005,Blinnikov2006} to obtain synthetic observables, specifically light curves and expansion velocity. Our modeling framework is based on the following \textit{inlists}: \texttt{12M\_pre\_ms\_to\_core\_collapse} and \texttt{ccsn\_IIp}, with most parameters remaining unchanged. A detailed description of the parameter choices and their effects is discussed in detail in \citet{Farmer} and \citet{Paxton2018}; We also refer to \citet{2022ApJ...930...34T, 2023ApJ...954..155T} for additional detailed descriptions of the setup and some of the key parameters. As observed, low-energy explosions result in matter fallback after the explosion. This has been quantified in MESA using the binding-energy fallback scheme \citet{Paxton2019,2019ApJ...879....3GGOLD}. It quantifies late-time fallback during the shock-propagation phase. For modeling SN~2024abfl observables, we primarily focus on the zero-age main-sequence (ZAMS) mass,
mixing length ($\alpha_{MLT}$), explosion energy ($\rm E_{exp}$), and nickel mass ($\rm M_{Ni}$).  The progenitor models are exploded in \texttt{MESA} via a thermal energy injection to a specified total
explosion energy, and the ejecta evolution is followed to just before shock breakout following \citet{Paxton2018} as discussed in \citet{2022ApJ...930...34T, 2023ApJ...954..155T} making use of the \citet{2016ApJ...821...76D} implementation for mixing via the Rayleigh-Taylor Instability. The models are then handed off to \texttt{STELLA} when the shock reaches an overhead mass coordinate of 0.05~$\rm M_\odot$. 

\begin{table*}[hbt!]

\caption{Summary of key parameters estimated in different works for SN~2024abfl}
\centering
\begin{tabular}{cccc|cc|c} \hline
$\rm M_{Ni}$          & $\rm M_{ej}$         &Distance    & $E(B-V)_{Total}$ & $\rm M_{ZAMS}$ & Methodology & Ref.      \\ 
$\rm(M_\odot)$ & $\rm(M_\odot)$ & (Mpc) & (mag) & $\rm(M_\odot)$ & & \\
\hline 
- & - & $15.6^{+6.1}_{-3.0}$ & 0.45 (CMD); 0.06 (Na ID) & 9-12 & Pre-SN Imaging & 1 \\
$0.01\pm0.001$ & 7.3 & $15.6^{+6.1}_{-3.0}$ &$0.28\pm0.11$ & 9 & 1-D SNEC; Nebular Spectrum & 2 \\
0.009 & $8.3^{+5.9}_{-2.8}$ & $15.6^{+6.1}_{-3.0}$ & $0.22\pm0.04$ & $\leq 15$ & Nebular Spectrum & 3 \\
$0.003\pm0.001$ & - & $8.73-9.9$~$\oplus$ & 0.19 & 8-12 & Pre-SN Imaging & 4 \\
0.003-0.004 & 7-8 & 7-9~$\oplus$ & 0.17 & 9-10 & 1-D MESA+Stella & This work \\
      
\hline

\hline
\end{tabular} \\
{\scriptsize{ 1.\citet{2025ApJ...982L..55L}~2.\citet{2026arXiv260102638G}~3.\citet{2026arXiv260204309C}~4.\citet{2026arXiv260401806L}~$\oplus$~Measured independently using SCM and EPM.}}
\label{tab:allsn_param}
\end{table*}

We start our modeling efforts by evolving a typical 12 and 15~$\rm M_\odot$ RSG star and exploding it with $\rm \sim10^{50}~erg$ of energy. Table~\ref{tab:mesa_param} lists the various pre-SN properties of the evolved RSGs. When comparing the model luminosity and photospheric velocities with the observed values, we find significant discrepancies: the plateau luminosity and photospheric velocities are much higher in the model light curves than the observed values. Exploding them further with very low energies was not feasible, and most models fail to explode. Furthermore, these models show significant fallback during the shock propagation phase ($\rm \geq 1-2~M_\odot$) and require nickel to be mixed manually as the default nickel put about the core is being removed along with the fallback material. Hence, we did not pursue these models further and evolved models at the lower boundary of the constraints \citep[9-10~$\rm M_\odot$,][]{2025ApJ...982L..55L}. We could explode these models with much lower explosion energies ($\rm 0.01-0.1\times 10^{51}~erg$) and with minimal fallback ($\rm 0.20-0.50~M_\odot$). Some of the resulting model light curves and photospheric velocities are plotted in Figure~\ref{fig:mesa}. A visual inspection indicates that the 9-10~$\rm M_\odot$ ZAMS models trace the observed trend much better. Hence, we further fine-tune the 9-10~$\rm M_\odot$ models to ascertain the progenitor properties. The nickel mass is constrained between 0.003 - 0.004~$\rm M_\odot$, corroborating the previous estimates. We note that we did not achieve a comprehensive one-to-one match with the observations, but we did obtain enough insights to place meaningful constraints on the progenitor and explosion parameters. We find that SN~2024abfl could be explained by the explosion of a 9-10~$\rm M_\odot$ RSG star with $\rm E_{exp}\sim 0.01-0.05~foe$ and radii plausibly $<$~500~$\rm R_\odot$. However, the plateau luminosity is not yet fully reproduced and might require more compact progenitors or a different amount of $\rm ^{56}Ni$ mixing, which we have not explored here. Nevertheless, the light curve behavior is satisfactorily reproduced by a typical Fe-core collapse. These values do not deviate much from those obtained by \citep{2026arXiv260102638G} using SNEC modeling at a slightly larger distance. Even if we include the bolometric light curve obtained using a higher distance, it still falls within the modeled light curves obtained using 9-10~$\rm M_\odot$ progenitors. Hence, for both distances, the progenitor turns out to be a low-mass RSG exploded via Fe core collapse. We also exploded several KEPLER models in SNEC for the lower distance used in our case (see Appendix~\ref{appendix:1}), but they did not fit well even for very low explosion energies. At the edge of the CCSNe boundary, there is possibility of explosion having the electron capture origins resulting in an electron capture SN (ECSN). We can use a distance-independent probe to check whether there is a possibility of ECSN for SN~2024abfl. We use color relation derived in \citet{2024ApJ...970..163S}. For ECSN, $(g-r)_{t_{PT}/2}< 0.008 \times t_{PT}- 0.4,$ where $t_{PT}$ is the mid-point where the SN light curve transitions from plateau to the nickel tail. For, SN~2024abfl $(g-r)_{t_{PT}/2}\approx0.8\pm0.1~\rm mag$ which is greater than $0.008 \times t_{PT}- 0.4$ (0.6~mag). This diagnostic shows that SN~2024abfl origins are plausibly from a Fe core-collapse.  For comparison, we also plot electron capture SN (ECSN) models from \citet{2021MNRAS.503..797K} in Figure~\ref{fig:mesa}. Clearly, they show a much steeper, longer fall from the plateau to the tail owing to the low level of radioactive nickel, which is different from what we observe in SN~2024abfl.

\section{Summary}
\label{sec:end}

In this work, we presented detailed photometric and spectroscopic observations of the low-luminosity Type IIP supernova SN~2024abfl. Its explosion parameters and properties of the progenitor star were inferred using hydrodynamical modeling. We also obtained independent distance estimates to the SN host. SN~2024abfl showed a remarkably faint plateau with mid-plateau absolute magnitude of $\rm M_V\approx-13.8~mag$, putting it among the faintest Type IIP SNe known till now. One of the most intriguing properties of SN~2024abfl is its extremely flat plateau phase (0.1 mag/100 d in $V$-band), which is among the slowest ever measured for a large sample of Type II SNe. The spectral evolution is broadly consistent with that of other low-luminosity Type IIP SNe, showing narrow P-Cygni profiles and relatively low expansion velocities. Using MESA+STELLA modeling, we explored a range of progenitor models and explosion energies. Higher-mass progenitors ($\rm \gtrsim 12–15\,M_\odot$) were unable to reproduce the observed faint plateau luminosity and low expansion velocities. Models with lower-mass progenitors provide a better match to the observations. Our modeling suggests that SN~2024abfl likely originated from a low-energy (0.01-0.05~foe) explosion of a $\rm 9–10\,M_\odot$ red supergiant progenitor with a radius ($\rm <500\,R_\odot$) and $\rm 7-8~M_\odot$ of ejecta mass. We estimated $\rm ^{56}Ni$ mass to be $0.0035\pm0.006~M_\odot$. Table~\ref{tab:allsn_param} provides some of the parameters obtained in various other works on SN~2024abfl. Further detailed modeling, including multi-dimensional simulations and improved constraints on the host galaxy distance, will be required to better understand the explosion mechanism and to determine whether SN~2024abfl belongs to the same physical class as other low-luminosity Type IIP events or represents an even more extreme case within this population.

\begin{acknowledgments}
We thank the anonymous reviewer for constructive feedback and suggestions that improved the overall work.

GCA thanks the Indian National Science Academy for support under the INSA Senior Scientist Programme.

AS acknowledges support from the Knut and Alice Wallenberg Foundation through the ``Gravity Meets Light" project.

DKS and HD acknowledge support from the the Science \& Engineering Research Board (SERB), Anusandhan National Research Foundation (ANRF), Government of India, through the Core Research Grant project file No. CRG/2022/007688.

MS acknowledges financial support provided under the National Post Doctoral Fellowship (N-PDF; File Number: PDF/2023/002244) by the Science \& Engineering Research Board (SERB), Anusandhan National Research Foundation (ANRF), Government of India.

We thank the staff of IAO, Hanle, CREST, and Hosakote, who made these observations possible. The facilities at IAO and CREST are operated by the Indian Institute of Astrophysics, Bangalore.

The GROWTH India Telescope (GIT) is a 70-cm telescope with a 0.7-degree field of view, set up by the Indian Institute of Astrophysics (IIA) and the Indian Institute of Technology Bombay (IITB) with funding from Indo-US Science and Technology Forum and the Science and Engineering Research Board, Department of Science and Technology, Government of India. It is located at the Indian Astronomical Observatory (IAO, Hanle). We acknowledge funding by the IITB alumni batch of 1994, which partially supports the operation of the telescope.

This work has made use of data from the Asteroid Terrestrial-impact Last Alert System (ATLAS) project. The Asteroid Terrestrial-impact Last Alert System (ATLAS) project is primarily funded to search for near earth asteroids through NASA grants NN12AR55G, 80NSSC18K0284, and 80NSSC18K1575; byproducts of the NEO search include images and catalogs from the survey area. This work was partially funded by Kepler/K2 grant J1944/80NSSC19K0112 and HST GO-15889, and STFC grants ST/T000198/1 and ST/S006109/1. The ATLAS science products have been made possible through the contributions of the University of Hawaii Institute for Astronomy, the Queen’s University Belfast, the Space Telescope Science Institute, the South African Astronomical Observatory, and The Millennium Institute of Astrophysics (MAS), Chile.

This research has made use of the NASA/IPAC Extragalactic Database (NED), which is funded by the National Aeronautics and Space Administration and operated by the California Institute of Technology.

\end{acknowledgments}


%
%


\facilities{HCT:2.0m, GIT:0.7m, ATLAS, ZTF, Swift(UVOT)}

\software{astropy \citep{astropy:2013, astropy:2018, astropy:2022},
ds9 \citep{2000ascl.soft03002S},
emcee \citep{2013PASP..125..306F},
IRAF \citep{1993ASPC...52..173T},
Jupyter-notebook \citep{jupyter},
matplotlib \citep{Hunter:2007},
MESA,
numpy \citep{harris2020array},
pandas \citep{mckinney-proc-scipy-2010, reback2020pandas},
plot\_atlas\_fp.py \citep{Youngfp} 
scipy \citep{2020SciPy-NMeth},
STELLA,
Superbol \citep{mnicholl_2018_2155821}}

\appendix

\section{Data}
\label{app:data}

This section includes the apparent magnitudes estimated for SN~2024abfl from $Swift$/UVOT, GIT and HCT telescopes. The magnitudes are shown in Tables~\ref{tab:uvotphot},\ref{tab:hctphot} and \ref{tab:gitphot}.

\begin{table}[hbt!]
\centering
\caption{Photometry values obtained from the Swift/UVOT (ABmag).} 
\begin{tabular}{|c|c|cc|cc|cc|cc|cc|cc|} \hline
\multicolumn{2}{|c|}{Filters$~\rightarrow$} &    \multicolumn{2}{c|}{UVW2} & \multicolumn{2}{c|}{UVM2} & \multicolumn{2}{c|}{UVW1} &
\multicolumn{2}{c|}{U} & \multicolumn{2}{c|}{B} &
\multicolumn{2}{c|}{V} \\  
\hline
JD$^+$	&	Phase$^\ast$	&	mag	&	err	&	mag	&	err	&	mag	&	err	&	mag	&	err	&	mag	&	err	&	mag	&	err \\
\hline
635.4	&	6.4	&	19.63	&	0.2	&	19.25	&	0.19	&	18.67	&	0.16	&	17.04	&	0.11	&	16.98	&	0.15	&	16.62	&	0.2 \\
636.5	&	7.5	&	19.87	&	0.16	&	20.24	&	0.22	&	18.96	&	0.14	&	17.4	&	0.1	&	16.89	&	0.1	&	16.91	&	0.2 \\
639.4	&	10.4	&	21.18	&	0.27	&	22.02	&	0.62	&	20.06	&	0.19	&	18.15	&	0.1	&	17.09	&	0.08	&	16.88	&	0.13 \\
641.1	&	12.1	&	21.1	&	0.24	&	21.25	&	0.34	&	20.21	&	0.19	&	18.16	&	0.1	&	17.23	&	0.08	&	16.88	&	0.11 \\
645.7	&	16.7	&	21.63	&	0.35	&	21.6	&	0.38	&	20.4	&	0.24	&	18.97	&	0.17	&	17.47	&	0.1	&	16.8	&	0.12 \\
649.4	&	20.4	&	21.55	&	0.32	&	22.3	&	0.55	&	21	&	0.35	&	19.53	&	0.24	&	17.59	&	0.11	&	17.22	&	0.15\\

\hline
\end{tabular} \\
$^+$ 2460000.0$+$; $^{\ast}$ With reference to the explosion date (JD~2460629). 
\label{tab:uvotphot}
\end{table}

\begin{table}[hbt!]
\centering
\caption{Photometry values obtained from the 2.0m HCT (ABmag).} 
\begin{tabular}{|c|c|cc|cc|cc|cc|} \hline
\multicolumn{2}{|c|}{Filters$~\rightarrow$} &    \multicolumn{2}{c|}{g} & \multicolumn{2}{c|}{i} & \multicolumn{2}{c|}{r} &
\multicolumn{2}{c|}{z}\\  
\hline

JD$^+$	&	Phase$^\ast$	&	mag	&	err	&	mag	&	err	&	mag	&	err	&	mag	&	err	\\
\hline
630.5	&	1.5	&	17.06	&	0.04	&	17.26	&	0.05	&	17.10	&	0.04	&	-	&	-	\\
631.2	&	2.2	&	17.04	&	0.03	&	17.17	&	0.04	&	17.06	&	0.03	&	17.27	&	0.05	\\
632.3	&	3.3	&	17.04	&	0.10	&	17.17	&	0.05	&	17.04	&	0.04	&	17.35	&	0.14	\\
633.4	&	4.4	&	17.08	&	0.06	&	17.04	&	0.11	&	16.99	&	0.11	&	-	&	-	\\
656.2	&	27.2	&	17.52	&	0.32	&	16.88	&	0.04	&	16.99	&	0.02	&	16.89	&	0.06	\\
690.2	&	61.2	&	17.85	&	0.03	&	16.64	&	0.03	&	16.85	&	0.03	&	16.55	&	0.03	\\
692.3	&	63.3	&	17.96	&	0.04	&	16.68	&	0.07	&	16.92	&	0.04	&	-	&	- \\
740.1	&	111.1	&	17.96	&	0.02	&	16.68	&	0.06	&	16.90	&	0.02	&	16.62	&	0.08	\\
781.3	&	152.3	&	21.16	&	0.26	&	18.86	&	0.05	&	19.31	&	0.07	&	18.50	&	0.06	\\
785.3	&	156.3	&	-	&	-	&	18.87	&	0.05	&	-	&	-	&	-	&	-	\\
807.1	&	178.1	&	21.06	&	0.29	&	19.08	&	0.05	&	19.55	&	0.07	&	18.54	&	0.08	\\

\hline
\end{tabular} \\
$^+$ 2460000.0$+$; $^{\ast}$ With reference to the explosion date (JD~2460629). 
\label{tab:hctphot}
\end{table}

\startlongtable
\begin{deluxetable*}{|cccc|cccc|cccc|cccc|}
\tabletypesize{ \scriptsize}
\tablecaption{Photometry values obtained from the 0.7m GIT (ABmag).}
\tablehead{
\colhead{JD$^+$}	&	\colhead{Phase$^\ast$}&\colhead{g-mag}&\colhead{err}&	\colhead{JD$^+$}	&	\colhead{Phase$^\ast$}	&	\colhead{i-mag}	&	\colhead{err}	&	\colhead{JD$^+$}	&	\colhead{Phase$^\ast$}	&	\colhead{r-mag}	&	\colhead{err}	&	\colhead{JD$^+$	}&	\colhead{Phase$^\ast$}	&	\colhead{z-mag}	&	\colhead{err} }

\startdata
636.4	&	7.4	&	17.1	&	0.04	&	636.4	&	7.4	&	17	&	0.04	&	636.4	&	7.4	&	16.95	&	0.04	&	636.4	&	7.4	&	17	&	0.1	\\
637.1	&	8.1	&	17.13	&	0.04	&	637.1	&	8.1	&	17	&	0.06	&	637.1	&	8.1	&	16.98	&	0.04	&	639.1	&	10.1	&	17.04	&	0.1	\\
639.1	&	10.1	&	17.16	&	0.04	&	639.1	&	10.1	&	17.27	&	0.12	&	639.1	&	10.1	&	16.86	&	0.05	&	640.1	&	11.1	&	17.24	&	0.12	\\
640.1	&	11.1	&	17.17	&	0.03	&	640.1	&	11.1	&	16.99	&	0.04	&	640.1	&	11.1	&	16.91	&	0.04	&	643.3	&	14.3	&	16.44	&	0.23	\\
642.1	&	13.1	&	18.5	&	0.82	&	647.3	&	18.3	&	16.74	&	0.12	&	644.1	&	15.1	&	17.07	&	0.09	&	644.1	&	15.1	&	17.01	&	0.21	\\
644.1	&	15.1	&	17.71	&	0.18	&	648.2	&	19.2	&	16.86	&	0.08	&	648.2	&	19.2	&	17.22	&	0.04	&	647.3	&	18.3	&	16.63	&	0.1	\\
645.2	&	16.2	&	17.43	&	0.04	&	655.1	&	26.1	&	16.84	&	0.1	&	651.2	&	22.2	&	16.94	&	0.13	&	650.1	&	21.1	&	16.81	&	0.12	\\
655.1	&	26.1	&	17.47	&	0.07	&	658.2	&	29.2	&	16.81	&	0.06	&	655.1	&	26.1	&	16.9	&	0.1	&	655.1	&	26.1	&	16.85	&	0.11	\\
658.2	&	29.2	&	17.5	&	0.06	&	659.2	&	30.2	&	16.75	&	0.06	&	656.5	&	27.5	&	16.85	&	0.09	&	658.2	&	29.2	&	16.72	&	0.1	\\
659.2	&	30.2	&	17.45	&	0.07	&	660.3	&	31.3	&	16.75	&	0.43	&	658.2	&	29.2	&	16.91	&	0.05	&	659.2	&	30.2	&	16.72	&	0.09	\\
660.3	&	31.3	&	17.54	&	0.2	&	662.4	&	33.4	&	16.8	&	0.09	&	659.2	&	30.2	&	16.91	&	0.05	&	662.3	&	33.3	&	16.67	&	0.13	\\
662.3	&	33.3	&	17.58	&	0.09	&	666.2	&	37.2	&	16.68	&	0.05	&	660.3	&	31.3	&	16.94	&	0.22	&	666.2	&	37.2	&	16.69	&	0.07	\\
666.2	&	37.2	&	17.61	&	0.05	&	668.3	&	39.3	&	16.68	&	0.05	&	662.4	&	33.4	&	16.88	&	0.07	&	668.3	&	39.3	&	16.69	&	0.07	\\
668.3	&	39.3	&	17.58	&	0.05	&	669.2	&	40.2	&	16.64	&	0.04	&	666.2	&	37.2	&	16.89	&	0.06	&	669.2	&	40.2	&	16.57	&	0.08	\\
669.2	&	40.2	&	17.57	&	0.06	&	671.1	&	42.1	&	16.66	&	0.06	&	669.2	&	40.2	&	16.85	&	0.03	&	671.1	&	42.1	&	16.56	&	0.08	\\
671.2	&	42.2	&	17.62	&	0.03	&	673.3	&	44.3	&	16.57	&	0.09	&	671.2	&	42.2	&	16.85	&	0.06	&	674.1	&	45.1	&	16.55	&	0.08	\\
673.3	&	44.3	&	17.67	&	0.08	&	674.1	&	45.1	&	16.64	&	0.05	&	672.3	&	43.3	&	16.83	&	0.1	&	675.2	&	46.2	&	16.57	&	0.08	\\
674.1	&	45.1	&	17.63	&	0.03	&	675.1	&	46.1	&	16.66	&	0.06	&	673.2	&	44.2	&	16.82	&	0.04	&	677.3	&	48.3	&	16.49	&	0.08	\\
675.1	&	46.1	&	17.62	&	0.03	&	677.3	&	48.3	&	16.59	&	0.09	&	674.1	&	45.1	&	16.85	&	0.03	&	683.1	&	54.1	&	16.15	&	0.1	\\
677.3	&	48.3	&	17.67	&	0.04	&	683.1	&	54.1	&	16.63	&	0.48	&	675.1	&	46.1	&	16.84	&	0.03	&	690.2	&	61.2	&	16.43	&	0.09	\\
683.1	&	54.1	&	17.65	&	0.11	&	690.2	&	61.2	&	16.61	&	0.19	&	677.3	&	48.3	&	16.86	&	0.05	&	691.2	&	62.2	&	16.59	&	0.08	\\
691.2	&	62.2	&	17.72	&	0.07	&	691.2	&	62.2	&	16.54	&	0.05	&	683.1	&	54.1	&	16.69	&	0.18	&	693.2	&	64.2	&	16.52	&	0.09	\\
693.2	&	64.2	&	17.77	&	0.05	&	693.2	&	64.2	&	16.57	&	0.04	&	690.2	&	61.2	&	16.81	&	0.05	&	696.4	&	67.4	&	13.9	&	0.01	\\
696.4	&	67.4	&	17.93	&	0.13	&	696.4	&	67.4	&	16.63	&	0.12	&	691.2	&	62.2	&	16.81	&	0.05	&	697.3	&	68.3	&	16.43	&	0.08	\\
697.3	&	68.3	&	17.83	&	0.05	&	697.3	&	68.3	&	16.55	&	0.06	&	693.2	&	64.2	&	16.82	&	0.03	&	699.2	&	70.2	&	16.5	&	0.07	\\
700.2	&	71.2	&	17.78	&	0.05	&	699.2	&	70.2	&	16.51	&	0.05	&	696.4	&	67.4	&	16.81	&	0.05	&	700.2	&	71.2	&	16.57	&	0.07	\\
701.2	&	72.2	&	17.35	&	0.13	&	700.2	&	71.2	&	16.53	&	0.05	&	697.3	&	68.3	&	16.81	&	0.04	&	701.2	&	72.2	&	16.46	&	0.09	\\
702.2	&	73.2	&	17.85	&	0.05	&	701.2	&	72.2	&	16.59	&	0.05	&	699.2	&	70.2	&	16.8	&	0.04	&	702.2	&	73.2	&	16.37	&	0.06	\\
703.2	&	74.2	&	17.83	&	0.06	&	702.2	&	73.2	&	16.53	&	0.07	&	700.2	&	71.2	&	16.79	&	0.04	&	703.2	&	74.2	&	16.38	&	0.07	\\
706.2	&	77.2	&	17.89	&	0.06	&	703.2	&	74.2	&	16.51	&	0.05	&	701.2	&	72.2	&	16.8	&	0.04	&	706.2	&	77.2	&	16.56	&	0.1	\\
708.2	&	79.2	&	17.83	&	0.03	&	706.2	&	77.2	&	16.61	&	0.06	&	702.2	&	73.2	&	16.79	&	0.04	&	708.2	&	79.2	&	16.44	&	0.05	\\
710.2	&	81.2	&	17.79	&	0.04	&	708.2	&	79.2	&	16.46	&	0.06	&	703.2	&	74.2	&	16.81	&	0.04	&	710.2	&	81.2	&	16.31	&	0.07	\\
714.3	&	85.3	&	17.77	&	0.06	&	710.2	&	81.2	&	16.46	&	0.05	&	706.2	&	77.2	&	16.8	&	0.04	&	714.3	&	85.3	&	16.38	&	0.07	\\
716.3	&	87.3	&	17.79	&	0.09	&	714.3	&	85.3	&	16.48	&	0.04	&	708.2	&	79.2	&	16.78	&	0.06	&	716.3	&	87.3	&	16.42	&	0.08	\\
717.4	&	88.4	&	17.78	&	0.13	&	716.3	&	87.3	&	16.49	&	0.05	&	710.2	&	81.2	&	16.78	&	0.04	&	719.3	&	90.3	&	16.47	&	0.08	\\
719.3	&	90.3	&	17.8	&	0.13	&	719.3	&	90.3	&	16.47	&	0.08	&	714.3	&	85.3	&	17.25	&	0.14	&	720.4	&	91.4	&	16.36	&	0.1	\\
720.4	&	91.4	&	17.81	&	0.13	&	720.4	&	91.4	&	16.57	&	0.09	&	716.3	&	87.3	&	16.79	&	0.05	&	722.4	&	93.4	&	16.41	&	0.07	\\
722.4	&	93.4	&	17.77	&	0.08	&	722.4	&	93.4	&	16.53	&	0.05	&	717.4	&	88.4	&	16.73	&	0.05	&	723.3	&	94.3	&	16.61	&	0.08	\\
723.3	&	94.3	&	17.85	&	0.09	&	723.3	&	94.3	&	16.52	&	0.05	&	719.3	&	90.3	&	16.8	&	0.09	&	724.3	&	95.3	&	16.48	&	0.07	\\
724.3	&	95.3	&	17.76	&	0.06	&	724.3	&	95.3	&	16.59	&	0.05	&	720.4	&	91.4	&	16.77	&	0.08	&	730.3	&	101.3	&	16.52	&	0.09	\\
730.3	&	101.3	&	17.83	&	0.08	&	730.4	&	101.4	&	16.61	&	0.16	&	722.4	&	93.4	&	16.79	&	0.04	&	731.3	&	102.3	&	16.5	&	0.08	\\
731.3	&	102.3	&	17.81	&	0.05	&	731.3	&	102.3	&	16.56	&	0.06	&	723.3	&	94.3	&	16.78	&	0.05	&	737.3	&	108.3	&	16.53	&	0.11	\\
737.3	&	108.3	&	17.91	&	0.06	&	737.3	&	108.3	&	16.67	&	0.05	&	724.3	&	95.3	&	16.82	&	0.04	&	764.2	&	135.2	&	18.77	&	0.37	\\
743.4	&	114.4	&	17.9	&	0.2	&	743.4	&	114.4	&	16.39	&	0.17	&	730.4	&	101.4	&	16.84	&	0.04	&	764.2	&	135.2	&	18.13	&	0.28	\\
761.3	&	132.3	&	20.75	&	0.42	&	761.3	&	132.3	&	18.37	&	0.17	&	731.3	&	102.3	&	16.84	&	0.04	&	765.1	&	136.1	&	18.33	&	0.29	\\
764.2	&	135.2	&	20.99	&	0.38	&	764.2	&	135.2	&	18.64	&	0.13	&	737.3	&	108.3	&	16.85	&	0.05	&	765.1	&	136.1	&	18.4	&	0.3	\\
765.1	&	136.1	&	21.11	&	0.3	&	765.1	&	136.1	&	18.61	&	0.14	&	743.4	&	114.4	&	16.67	&	0.19	&	766.2	&	137.2	&	18.57	&	0.34	\\
766.2	&	137.2	&	20.81	&	0.31	&	766.2	&	137.2	&	18.78	&	0.15	&	755.3	&	126.3	&	18.6	&	0.12	&		&		&		&		\\
770.3	&	141.3	&	21.32	&	0.64	&	771.2	&	142.2	&	18.79	&	0.18	&	761.3	&	132.3	&	19.27	&	0.23	&		&		&		&		\\
771.2	&	142.2	&	20.75	&	0.48	&	772.2	&	143.2	&	18.33	&	0.22	&	764.2	&	135.2	&	19.2	&	0.27	&		&		&		&		\\
772.2	&	143.2	&	21.02	&	0.69	&		&		&		&		&	765.1	&	136.1	&	19.08	&	0.11	&		&		&		&		\\
	&		&		&		&		&		&		&		&	766.2	&	137.2	&	19.13	&	0.13	&		&		&		&		\\
	&		&		&		&		&		&		&		&	770.2	&	141.2	&	19.4	&	0.15	&		&		&		&		\\
	&		&		&		&		&		&		&		&	771.2	&	142.2	&	19.12	&	0.14	&		&		&		&		\\		
\hline
\multicolumn{16}{c}{\footnotesize {$^+$ 2460000.0$+$;
$^{\ast}$ With reference to the explosion date (JD~2460629). }}
\enddata
\label{tab:gitphot}
\end{deluxetable*}

\section{Supplementary material}
\label{appendix:1}
\begin{figure}[htb!]
    \centering
        
    \resizebox{0.7\hsize}{!}{\includegraphics{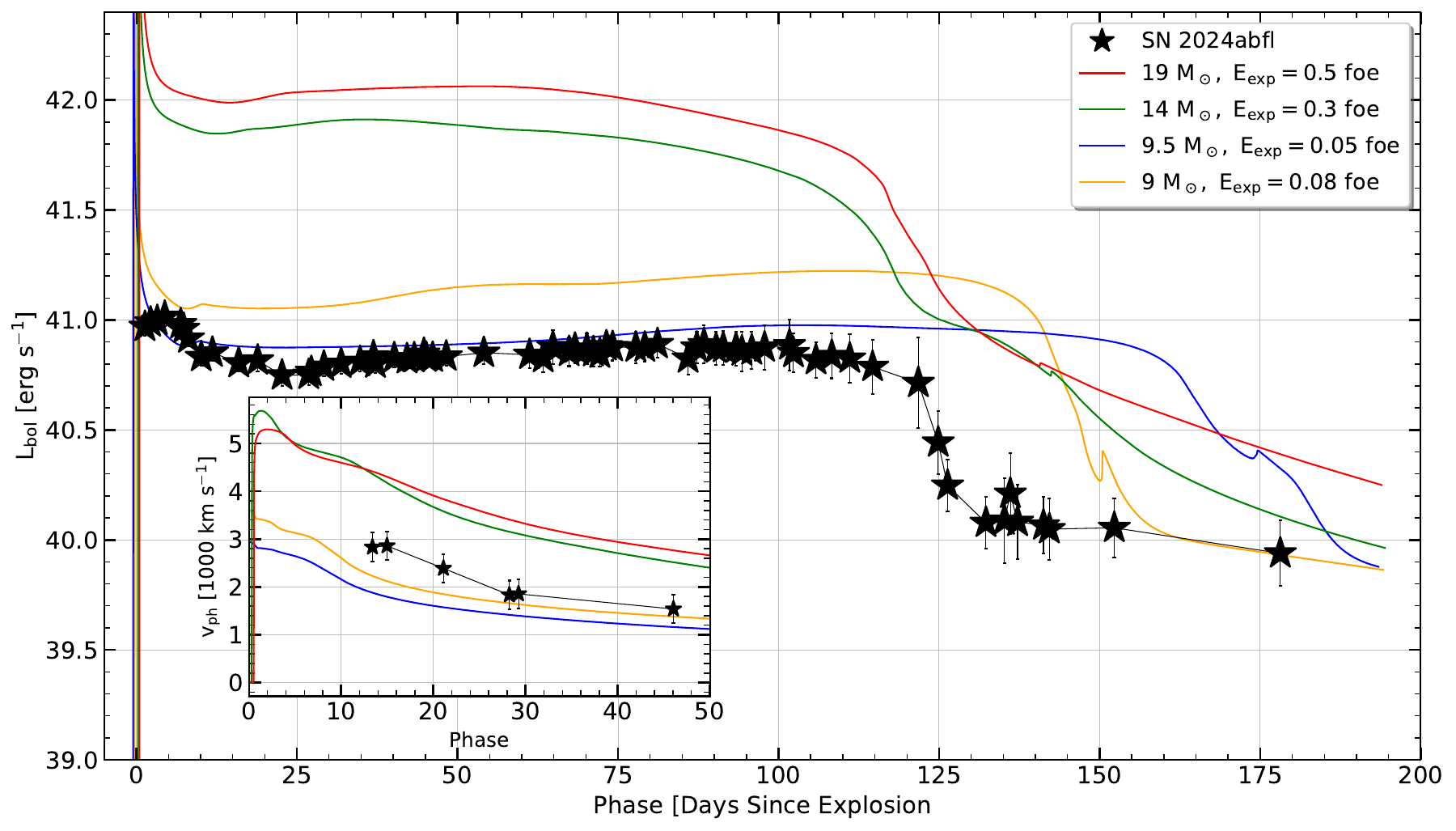}}

    \caption{Model light curves and photospheric velocities obtained for various KEPLER code progenitors in comparison to the observed values for SN~2024abfl. (We fixed $^{56}Ni=0.003~M_\odot$ for all the models) }
    \label{fig:snec}
\end{figure}

\citet{2026arXiv260102638G}, in their recent work, had obtained a similar value of plausible progenitor ($\rm (9~M_\odot$) using an existing set of comprehensive progenitor models obtained from the KEPLER code \citep{2016Sukhbold}. We had also utilized these evolved progenitors from \citet{2016Sukhbold} and exploded them using the SuperNova Explosion Code \citep[SNEC][]{2015ascl.soft05033M} to compare the model light curves with the bolometry obtained at 9.6~Mpc distance. SNEC is an open-source code that solves for Lagrangian hydrodynamics supplemented with radiative diffusion for the spherically expanding envelopes of CCSNe. It is also apt for hydrogen-rich SNe, as it accounts for both recombination and radioactive nickel. For a given progenitor model, input explosion energy and seed nickel mass can be provided to obtain bolometric and multiband light curves in the blackbody approximation. We exploded these with energies of the order $\rm \sim 10^{50}~erg$ or less. With much lower energies, these models fail to undergo an explosion. Model light curves and photospheric velocities derived from these also rule out massive-star models. We also test 9 and 9.5 $\rm M_\odot$ ZAMS models; these are the only models that closely match the observed trend in both velocities and luminosities, but not satisfactorily. However, the 9~$\rm M_\odot$ model can easily fit the observed bolometric light curve at a slightly higher distance, as seen in \citet{2026arXiv260102638G}, which we also obtained, albeit by a slightly different approach. We present some of these models in Figure~\ref{fig:snec}.

\begin{figure}[htb!]
    \centering
        
    \resizebox{0.6\hsize}{!}{\includegraphics{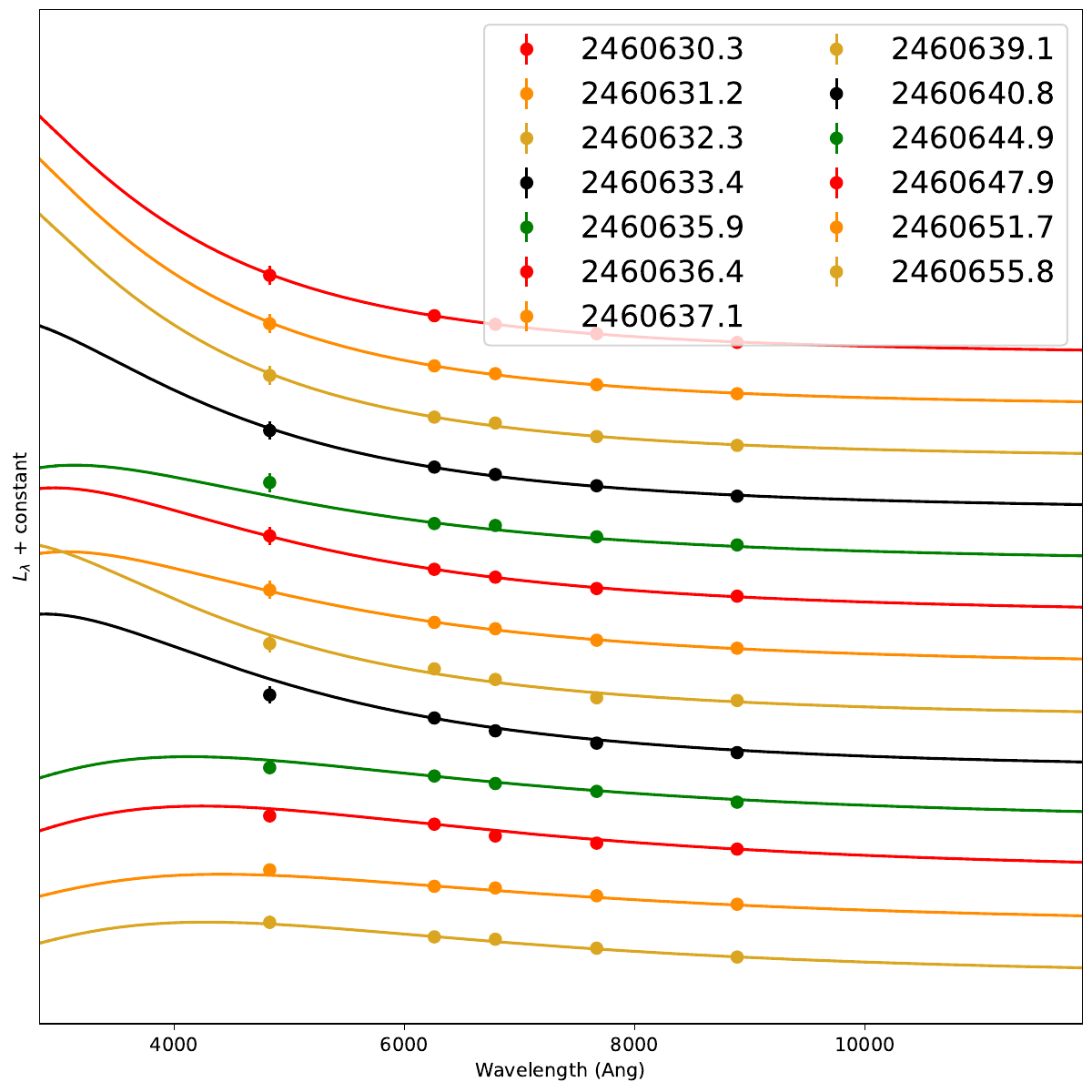}}

    \caption{\textbf{Blackbody fits to the optical data points obtained from the SuperBol.} }
    \label{fig:bbfitsoptical}
\end{figure}

Figure~\ref{fig:bbfitsoptical} shows the blackbody fits to the optical data for the initial few days. 

\bibliography{2024abfl}{}
\bibliographystyle{aasjournalv7}

\end{document}